\documentclass[namedreferences]{solarphysics}

\usepackage[hyperref,optionalrh]{spr-sola-addons} 
\usepackage{graphicx}        
\usepackage{amssymb}        
\usepackage{color}           
\usepackage{breakurl}        
\usepackage{caption}
\usepackage{lmodern}
\usepackage{textcomp}
\usepackage{microtype}
\usepackage{geometry}
\usepackage{enumerate}
\usepackage{url}

\graphicspath{{./}{figures/}}

\chardef\us=`\_

\begin{document}

\begin{article}
\begin{opening}

\title{Long-term Evolution of the Solar Corona Using PROBA2 Data\\ }

\author[addressref={aff1,aff2},corref,email={marilena.mierla@oma.be}]{\inits{M.}\fnm{Marilena}~\lnm{Mierla}\orcid{0000-0003-4105-7364}}
\author[addressref=aff1,email={jan.janssens@oma.be}]{\inits{J.}\fnm{Jan}~\lnm{Janssens}}
\author[addressref=aff1,email={elke.dhuys@oma.be}]{\inits{E.}\fnm{Elke}~\lnm{D'Huys}\orcid{0000-0002-2914-2040}}
\author[addressref=aff1,email={laurence.wauters@oma.be}]{\inits{L.}\fnm{Laurence}~\lnm{Wauters}}
\author[addressref=aff1,email={matthew.west@oma.be}]{\inits{M.J.}\fnm{Matthew J.}~\lnm{West}\orcid{0000-0002-0631-2393}}
\author[addressref=aff3,email={daniel.seaton@noaa.gov}]{\inits{D.B.}\fnm{Daniel B.}~\lnm{Seaton}\orcid{0000-0002-0494-2025}}
\author[addressref=aff1,email={david.berghmans@oma.be}]{\inits{D.}\fnm{David}~\lnm{Berghmans}\orcid{0000-0003-4052-9462}}
\author[addressref={aff4,aff1},email={elena.podladchikova@pmodwrc.ch}]{\inits{E.}\fnm{Elena}~\lnm{Podladchikova}\orcid{0000-0003-1679-0986}}

       
\address[id=aff1]{Solar-Terrestrial Centre of Excellence - SIDC, Royal Observatory of Belgium, Avenue Circulaire 3, Brussels 1180, Belgium}
\address[id=aff2]{Institute of Geodynamics of the Romanian Academy, Bucharest, Romania}
\address[id=aff3]{Cooperative Institute for Research in Envionmental Sciences, University of Colorado, Boulder, Colorado, U.S.A}
\address[id=aff4]{Physikalisch-Meteorologisches Observatorium Davos / World Radiation Center, Davos, Switzerland}

\runningauthor{M. Mierla \textit{et al.}}
\runningtitle{Long-Term Corona}

\begin{abstract}
We use \textit{The Sun Watcher with Active Pixel System detector and Image Processing} (SWAP) imager onboard \textit{the Project for Onboard Autonomy 2} (PROBA2) mission to study the evolution of large-scale EUV structures in the solar corona observed throughout Solar Cycle 24 (from 2010 to 2019). We discuss the evolution of the on-disk coronal features and at different heights above the solar surface based on EUV intensity changes. We also look at the evolution of the corona in equatorial and polar regions and compare them at different phases of the solar cycle, as well as with sunspot number evolution and with the PROBA2/\textit{Lyman-Alpha Radiometer} (LYRA) signal. The main results are as follows: The three time series (SWAP on-disk average brightness, sunspot number and LYRA irradiance) are very well correlated, with correlation coefficients around 0.9. The average rotation rate of bright features at latitudes of +15$^{\circ}$, 0$^{\circ}$, and -15$^{\circ}$ was around 15\,degree\,day$^{-1}$ throughout the period studied. A secondary peak in EUV averaged intensity at the Poles was observed on the descending phase of SC24. These peaks (at North and South poles respectively) seem to be associated with the start of the development of the (polar) coronal holes. Large-scale off-limb structures were visible from around March 2010 to around March 2016, meaning that they were absent at the minimum phase of solar activity. A fan at the North pole persisted for more than 11 Carrington rotations (February 2014 to March 2015), and it could be seen up to altitudes of 1.6\,R$_{\odot}$. 
\end{abstract}
\end{opening}

\section{Introduction}
     \label{S-Introduction} 

The aim of this work is to study the evolution of the solar corona between January 2010 and June 2019, by using data from the \textit{Sun Watcher with Active Pixels and Image Processing} (SWAP) EUV solar telescope \citep{Seaton2013a, Halain2013} and the \textit{Large Yield RAdiometer} (LYRA:  \citet{Dominique2013} onboard the \textit{Project for On Board Autonomy 2} (PROBA2) spacecraft \citep{Santandrea2013} as well as the International Sunspot Number (ISN) dataset.
The period covers the Carrington rotations (CR) from 2093 to 2218, corresponding to most of Solar Cycle 24 (SC24). Due to the large field of view of SWAP, it is possible to study the evolution of the EUV solar corona up to around 1.6 solar radii (R$_{\odot}$) around the edge of the image and 1.7\,R$_{\odot}$ in the corners of the image, throughout SC24. Large-scale features as streamers and pseudostreamers as well as coronal fans are easily identified in the field of view (FOV) of SWAP images \citep[see, \textit{e.g.},][]{Seaton2013b, Goryaev2014, Rachmeler2014, Talpeanu2016}. These features are worth studying, as they are thought to be the sources of the solar wind.

It is known that there are two main types of solar wind: a fast, more steady, solar wind coming from coronal
holes (CHs) and a slow, more variable, solar wind associated with streamers \citep{Gosling1981, Sheeley1997,
Strachan2002, McComas2008, Abbo2016}. An intermediate (from slow to fast) wind originates in and around pseudostreamers \citep{Wang2007, Riley2012, Wang2012, Panasenco2013}.

Both closed and open magnetic field structures are observed in these regions of the solar corona (from 1 to 1.7\,R$_{\odot}$). As it is difficult to measure directly the topology of the magnetic field in the solar corona, models are typically used. The most applied model is Potential Field Source Surface (PFSS) first introduced by \citet{Schatten1969, Altschuler1969}. The key topological element of such a model is the boundary between open and closed field lines, the so called source surface. Field lines piercing through the source surface are considered as open, while those forming closed loops below it are identified as belonging to closed magnetic structures. The height at which this surface is fixed by most of the researchers is 2.5\,R$_{\odot}$, but some recent studies have shown that lower heights (as low as 1.2\,R$_{\odot}$) better fit the observations \citep{Asvestari2019, Bale2019}.

Nevertheless there are also observational signatures of closed/open magnetic coronal structures, usually indicated in the solar EUV and white light images through features like loops, streamers, pseudostreamers, and coronal fans.

\vspace{0.3cm}

{\itshape Loops}

Coronal loops are closed field regions, defined by a series of magnetic flux tubes, or loop structures, piercing the atmosphere from below and highlighted by hot plasma trapped on them. They are visible in both EUV and X-ray images. Most quiet-Sun potential loops are seen forming lower down in the solar atmosphere. However, following an eruption it is possible for loop systems to grow due to the pile-up of magnetic structures from the reconnection region. An unusually large post-eruptive loop system was observed by SWAP on 14 October 2014, which extended out to heights exceeding 1.5\,R$_{\odot}$ \citep{West2015}. Similar observations of loop structures, observed in X-ray images, extending out to 1.6\,R$_{\odot}$ were reported by \citet{Svestka1997}. Active region loops, which are most prominently observed in EUV images, rarely exceed heights of around 1.15\,R$_{\odot}$ \citep[see the review by][]{Reale2014}.

\vspace{0.3cm}

{\itshape Streamers and Pseudostreamers}

As defined by \citet{Rachmeler2014}, a coronal streamer is a magnetic structure overlying a single (or an odd number of) polarity inversion lines (PILs) with closed loops in the lower corona and oppositely oriented open magnetic field in the upper corona, such that a current sheet is present between the two open field domains.

A pseudostreamer is a magnetic structure overlying two (or a multiple of) PILs such that above the closed field, two domains of open field of the same polarity come together and no current sheet is present.

The tops of both types of streamer are often referred to as 'cusps', and highlight the intersection between closed and open field lines. The streamer cusp regions are often located at heights around 2.5 - 3.5\,R$_{\odot}$ \citep[see, \textit{e.g.},][]{Eselevich2007}, while the cusps of pseudostreamers are often located below 2 R$_{\odot}$ \citep{Wang2007, Wang2012}.

\vspace{0.3cm}

{\itshape Coronal Fans}

Coronal fans are large scale semi-static structures which extend to extremely large distances above the solar surface, up to several solar radii \citep{Talpeanu2016}, and they were only clearly identified in the large field of view of SWAP. Fans have a very long lifetime, and can be observed for several solar rotations (rotating with the Sun). They are seemingly open features, connecting to the solar surface at footpoints and extending away from the Sun in the other direction. Near their footpoints they appear almost radial, but above that, they bend around large closed loops. They do not generally close around the loops, but stretch out into interplanetary space \citep{Talpeanu2016, Seaton2013a}.

\citet{Talpeanu2016} studied 15 fans in the period between March 2010 and July 2010, and between July 2012 and October 2014. The author showed that fans can be associated with both types of structures, helmet streamers and pseudostreamers. If the fan has a knee, most likely it overlies a pseudostreamer, due to the magnetic configuration of its base. If the fan does not have a knee, then it is more likely associated with a bipolar helmet streamer.

\vspace{0.3cm}

By studying the solar corona over a long period of time, one can see how these large structures evolve and how they are associated with other regions of the solar atmosphere (sunspots, active regions, coronal holes, etc.). Sunspots and active regions are usually associated with closed magnetic field lines while coronal holes are associated with open field lines \citep[see, \textit{e.g.}, reviews by][]{Cranmer2009, vanDrielGesztelyi2015}. The large FOV of SWAP allows us to track the evolution of the large scale structures at different heights, up to 1.6\,R$_{\odot}$ and beyond in the corners of the image.

The article is structured as follows: In Section 2 we describe how we process the data. Section 3 covers the data analysis and is split to several subsections: Subsection 3.1 deals with the long term evolution of the solar corona, both on-disk and off-limb, whereas Subsection 3.2 is about the variation of the average solar corona in time, in different regions on-disk and off-limb. Discussion and conclusions are presented in Sections 4 and 5 respectively.

\section{Data Processing} \label{sec:processing}
We used the following data sets for this analysis: custom-processed Level-1 SWAP 17.4\,nm images, Level-3 LYRA irradiance time series, and version 2.0 of the ISN dataset.

The PROBA2/SWAP 17.4\,nm level-1 images were calibrated from level-0 images using the \textsf{SolarSoft IDL p2sw\_prep.pro} routine. This software performs dark current subtraction, flat-field correction, and point-spread function (PSF) deconvolution and applies image corrections to ensure that the Sun is centered and rotated with its North Pole up in the image frame \citep{Seaton2013a, Seaton2013b}.
Using median stacking on level-1 images inside blocks of 100\,minutes of observation we constructed a new composite image by computing the median value of every pixel in this subset of images. By computing the median value rather than the mean, we suppressed random noise in the images, rejected one-time events such as cosmic-ray hits, and excluded short-duration dynamic events, yielding a set of high-signal-to-noise-ratio images, which show essentially only the corona's most stable features over time.
These images are then regrouped by CR intervals, so one can also check the evolution of the corona for each CR. The datasets can be found at \url{proba2.sidc.be/swap/data/carrington\_rotations/}.

The blocks of 100 minutes (roughly 25 images per block) were chosen because the spacecraft rolls every 25 minutes, to suppress residual anisotropy in the images due to the different spacecraft orientations.
Also, the observation window is short enough that the effects of large-scale coronal evolution and solar rotation are relatively small, whilst filtering out short-term noise in the data \citep[see][]{Seaton2013b}.

The PROBA2/LYRA instrument makes solar-irradiance measurements using, among others, a zirconium filter corresponding to a 6 to 20\,nm bandpass (with a contribution below 2\,nm). For this work we used the LYRA level-3 dataset, which is calibrated and averaged over one-minute intervals (see \url{proba2.sidc.be/data/LYRA}). The calibrated data were obtained using the \textsf{SolarSoft IDL lyra\_get\_data.pro} routine. The calibration includes the rescaling of the data to represent a constant Sun-spacecraft distance of 1 \,AU, the removal of instrument dark current, a degradation trend correction, and the conversion of the data into physical units [Watt\,m$^{-2}$] \citep{Dominique2013}. 

The international sunspot number (ISN; version 2.0) dataset was taken from the SILSO webpage: \url{sidc.oma.be/silso/} and smoothed using the Meeus formula \citep{Meeus1958}. This is a 13-month smoothing formula using weighted coefficients, providing a smoothed dataset with more pronounced minima and maxima.

Data of the solar polar magnetic field strength were obtained from the Wilcox Solar Observatory \citep[WSO: \url{wso.stanford.edu/}, ] []{Hoeksema1995, Svalgaard1978}, measuring the line-of-sight field between approximately 55$^\circ$ and the solar poles. As the varying solar B$_\mathrm{0}$-angle provides a different view on the solar poles throughout the year, the raw data have been adjusted by WSO using a 20\,nHZ low pass filter, effectively removing this yearly geometric projection effect. The raw hemispheric ten-day values as well as the adjusted values were used in this article, averaged per month.

\section{Long-Term Evolution of the Solar Corona} \label{sec:analysis}

\subsection{The Evolution of the Global Solar Corona Based on the EUV Intensity Changes\label{subsec:global}}

\begin{figure}
\centering
\includegraphics[width=0.49\textwidth]{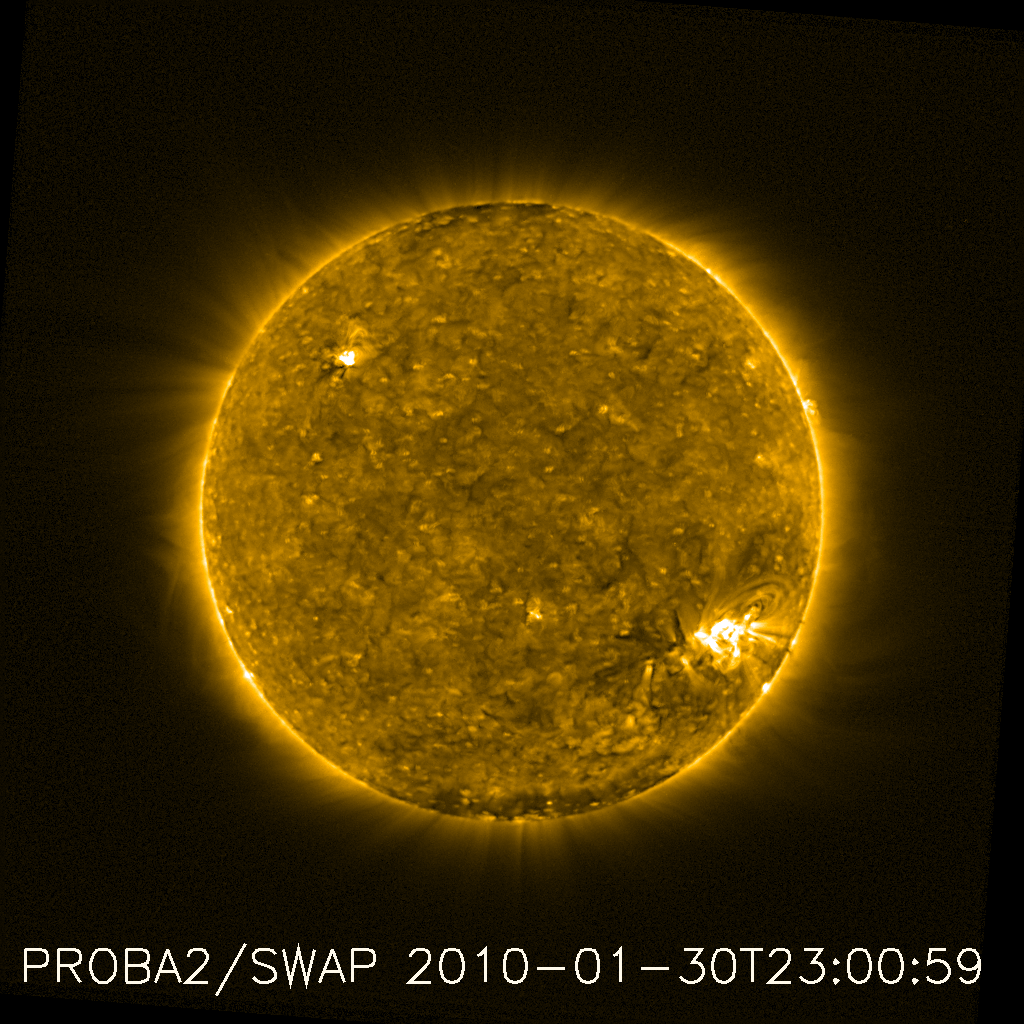}
\includegraphics[width=0.49\textwidth]{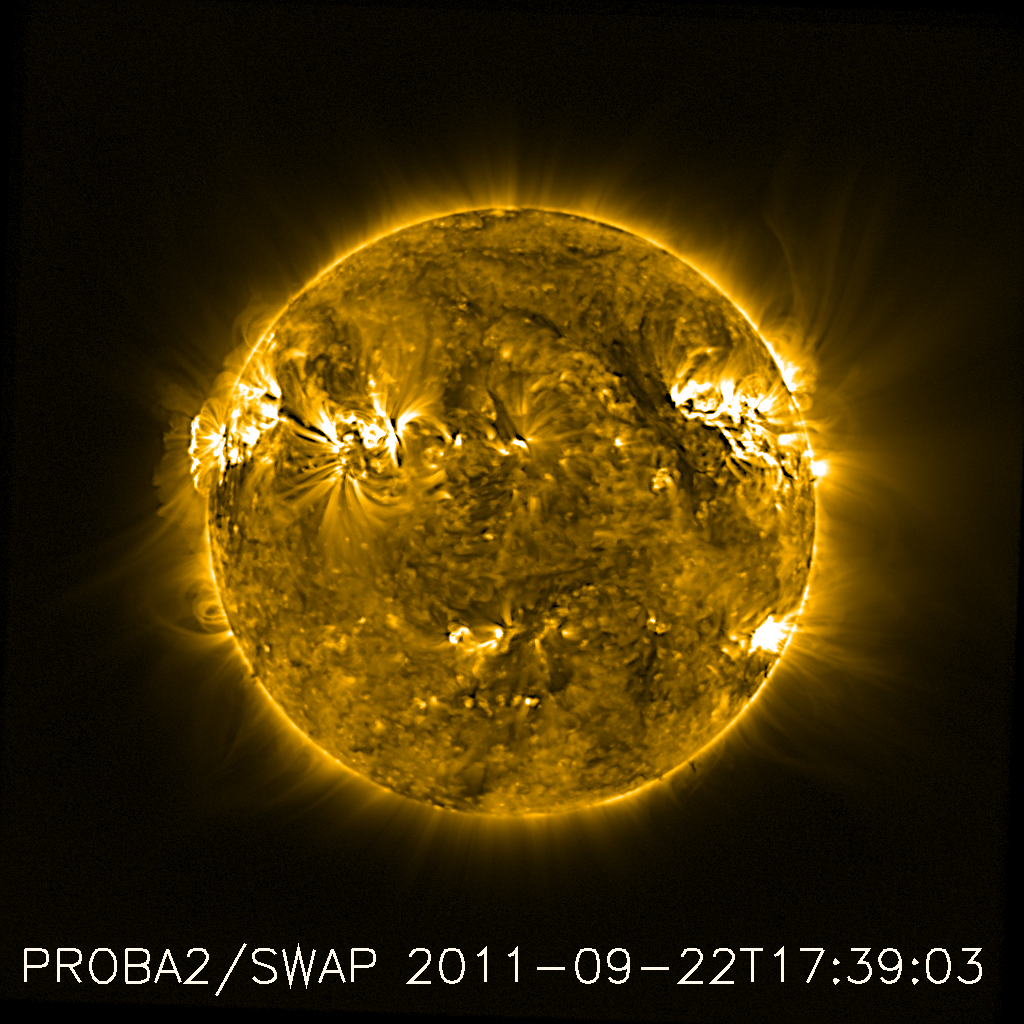}
\includegraphics[width=0.49\textwidth]{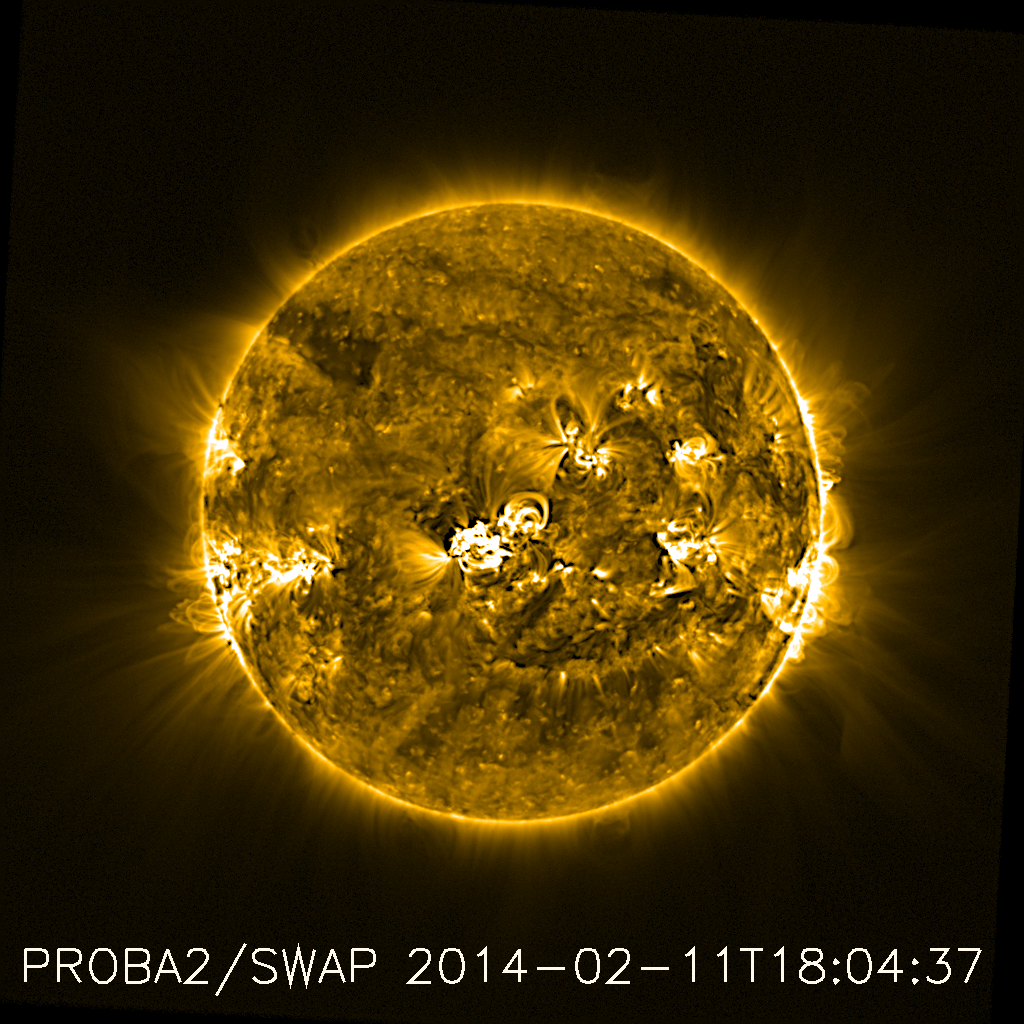}
\includegraphics[width=0.49\textwidth]{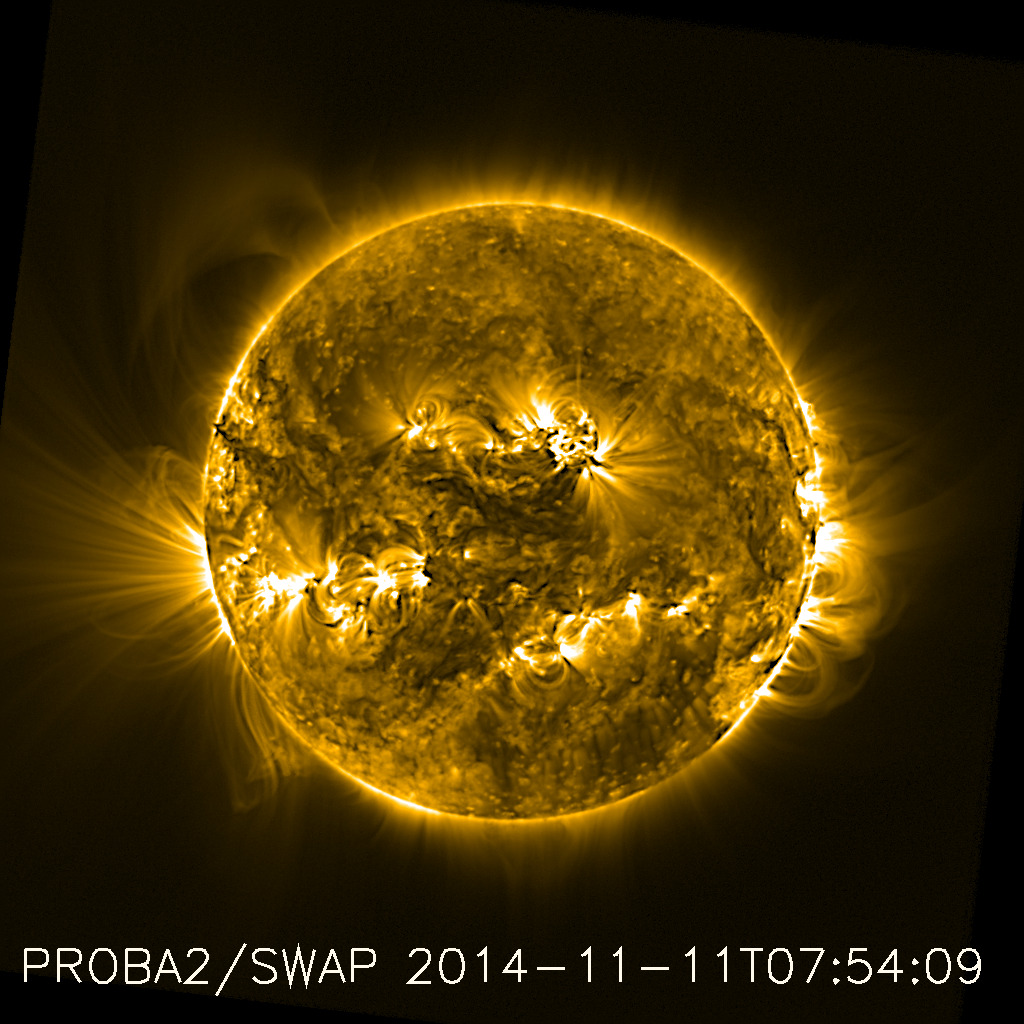}
\includegraphics[width=0.49\textwidth]{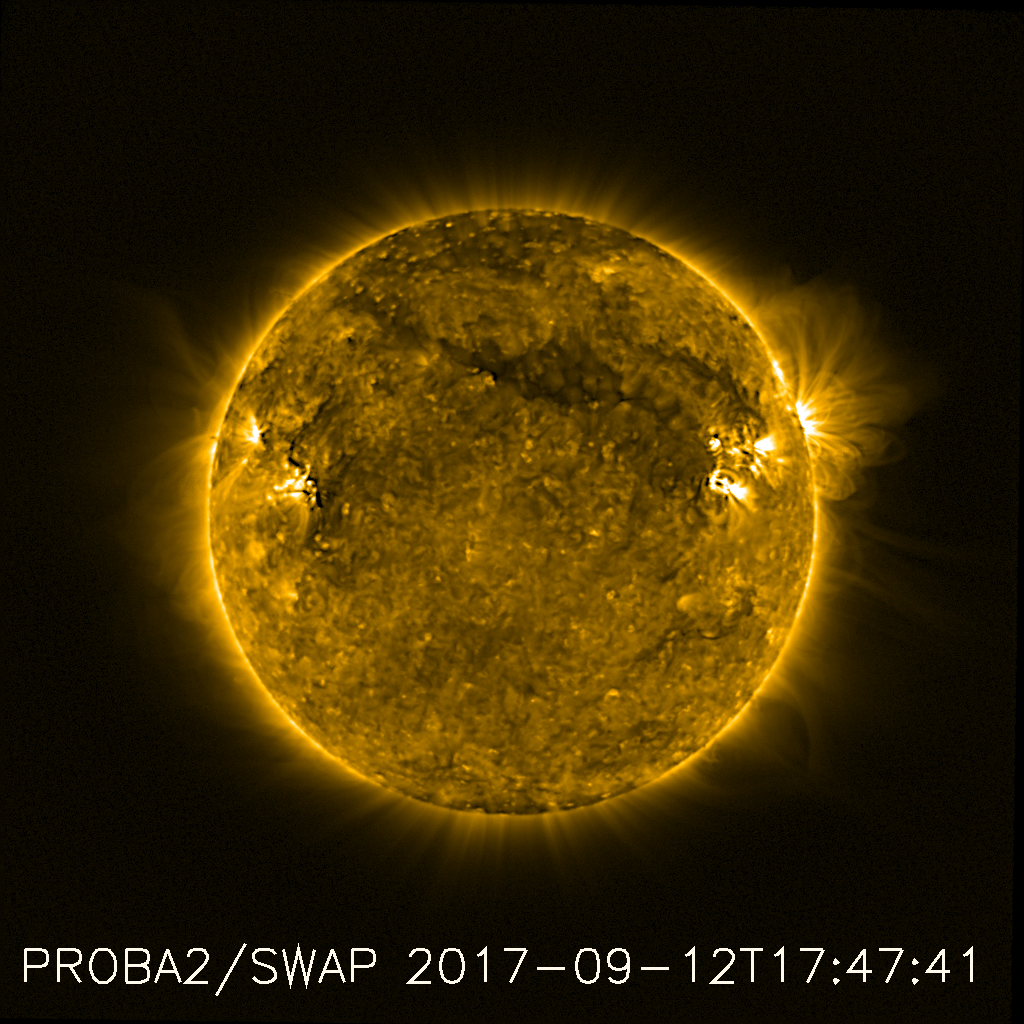}
\includegraphics[width=0.49\textwidth]{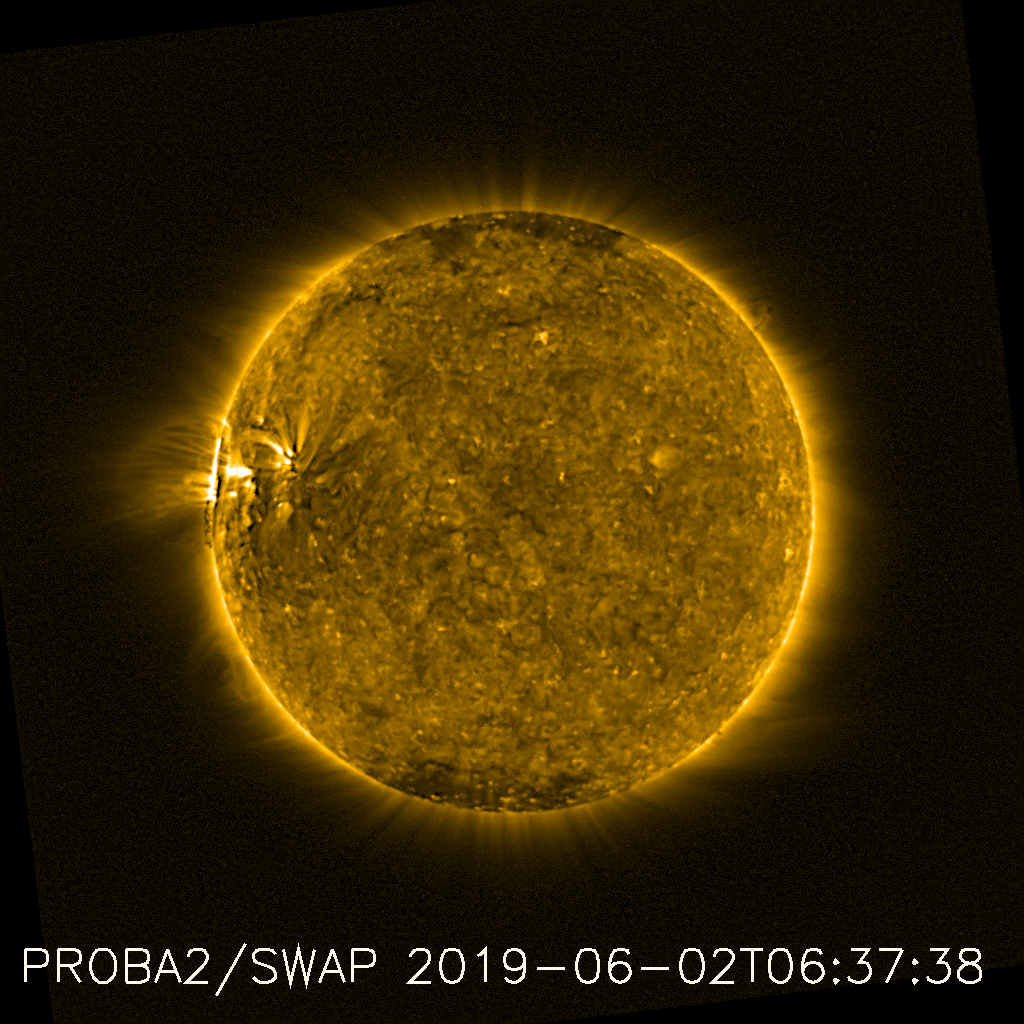}
\caption{SWAP stacked images (unsharp masked) at six moments of time showing the evolution of the solar corona from January 2010 to June 2019. The movie with one frame per Carrington rotation is available online.} \label{F-swapimg}
\end{figure}

Figure~\ref{F-swapimg} shows an overview of the evolution of the EUV solar corona as observed by SWAP from January 2010 to June 2019. By inspecting the corresponding movies obtained over this period, it was observed that a larger number of active regions (ARs) started to appear in February 2011 (in the northern hemisphere) and they became less frequent beginning in December 2016, reaching a very low number from September 2017 onward. Large-scale off-limb structures were visible from around March 2010 to around March 2016. 
This is roughly correlated with the evolution of the sunspot-activity cycle, which peaks late 2011 in the northern hemisphere and early 2014 in the southern hemisphere \citep[see also][]{Mordvinov2016}.

In the following sections, EUV intensity variations of on-disk and off-disk line-of-sight integrated emission of the solar corona are studied.

\subsubsection{The Evolution of the On-Disk Solar Corona}

\textit{SWAP Synoptic Maps}\\

\begin{figure}
\centering
\includegraphics[height=3cm,clip=]{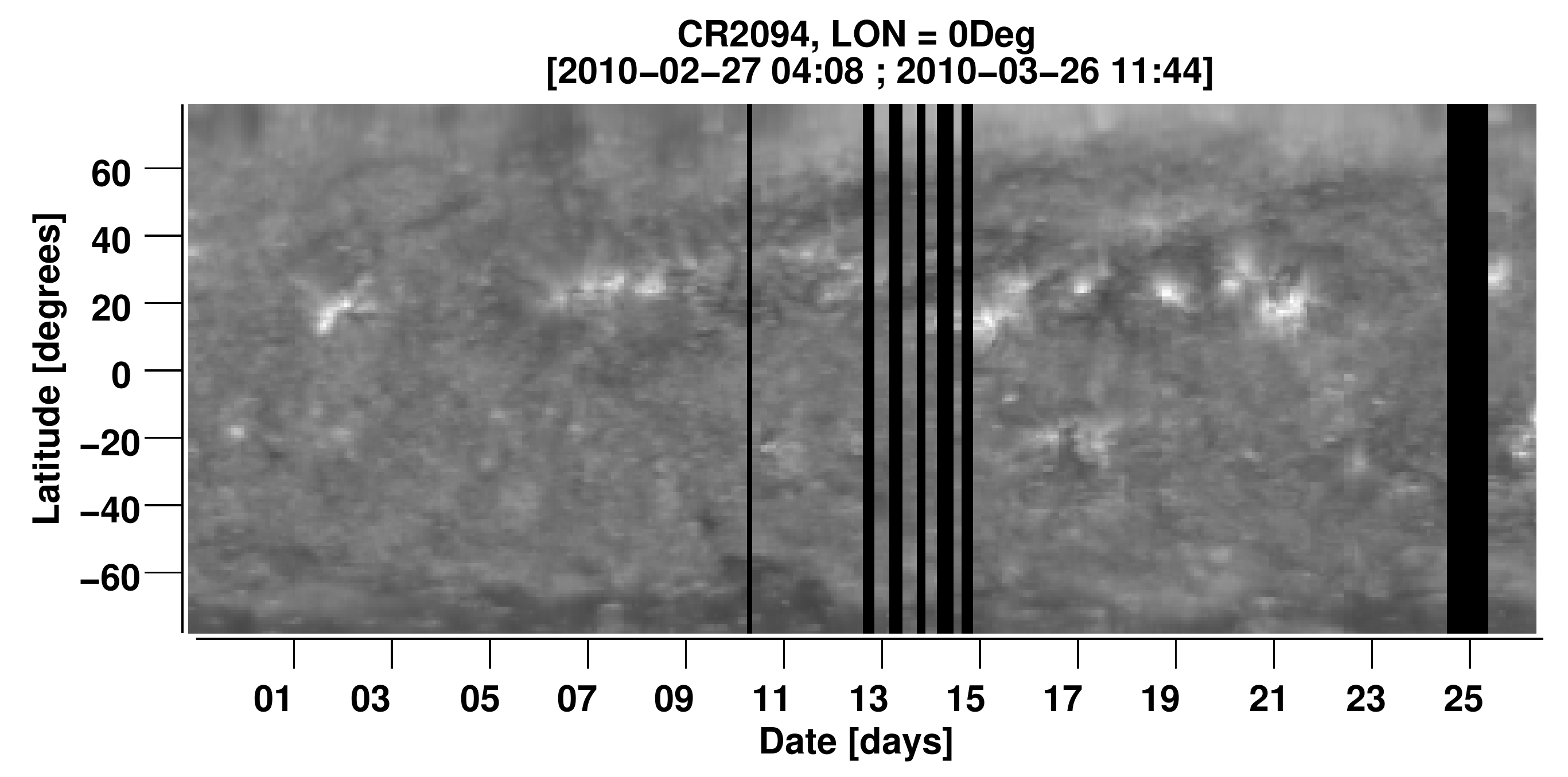}
\includegraphics[height=3cm,clip=]{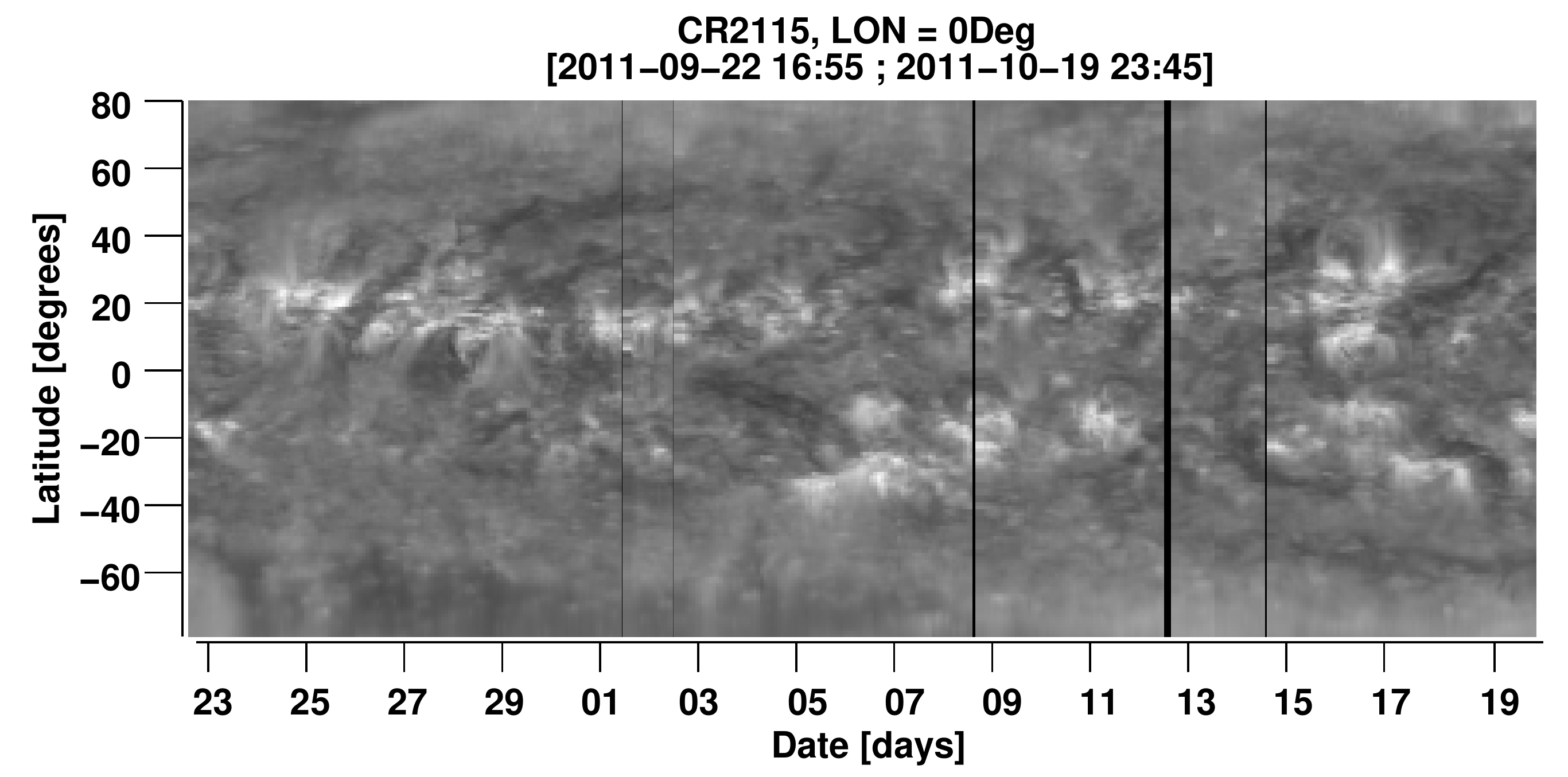}
\includegraphics[height=3cm,clip=]{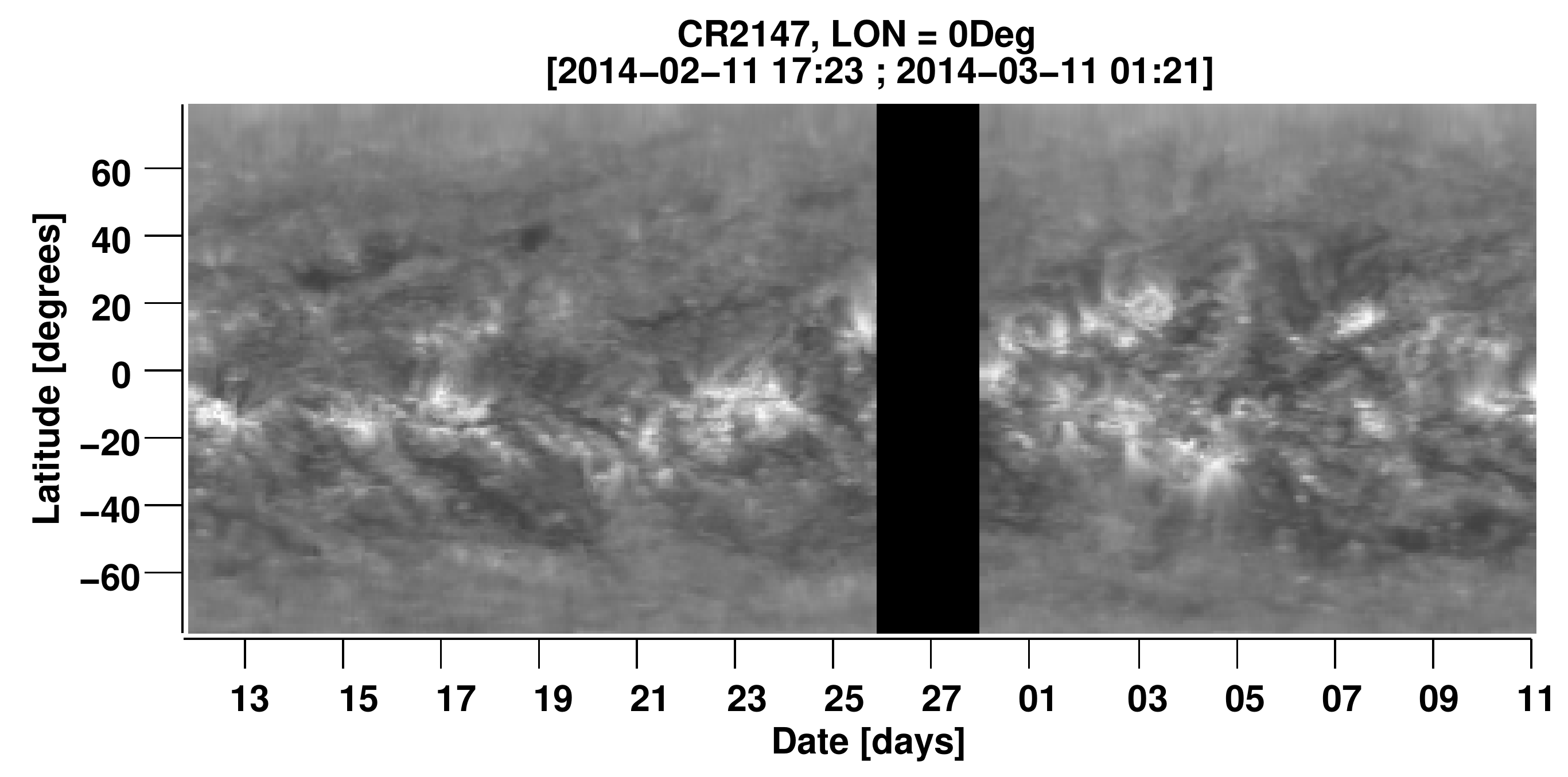}
\includegraphics[height=3cm,clip=]{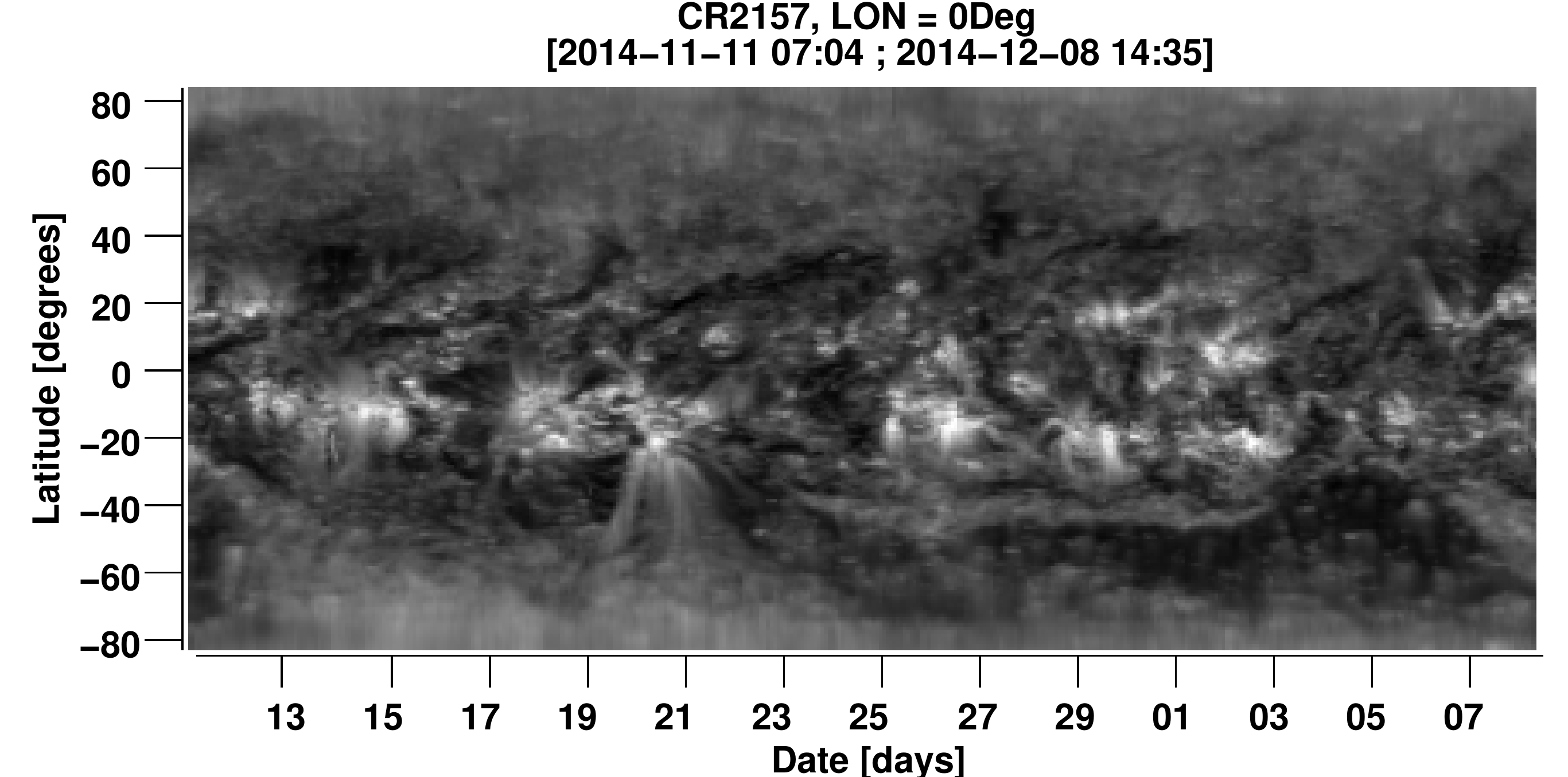}
\includegraphics[height=3cm,clip=]{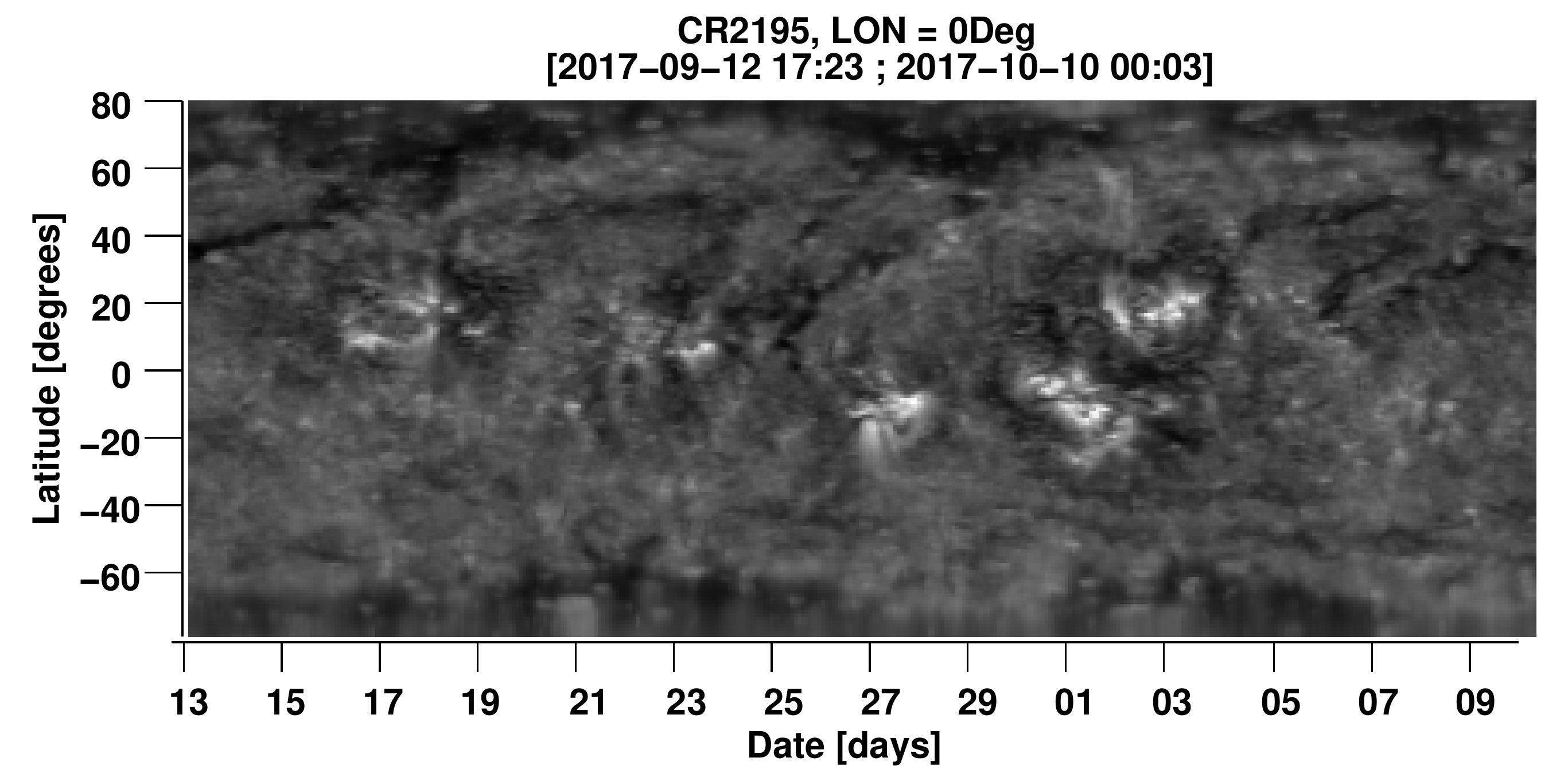}
\includegraphics[height=3cm,clip=]{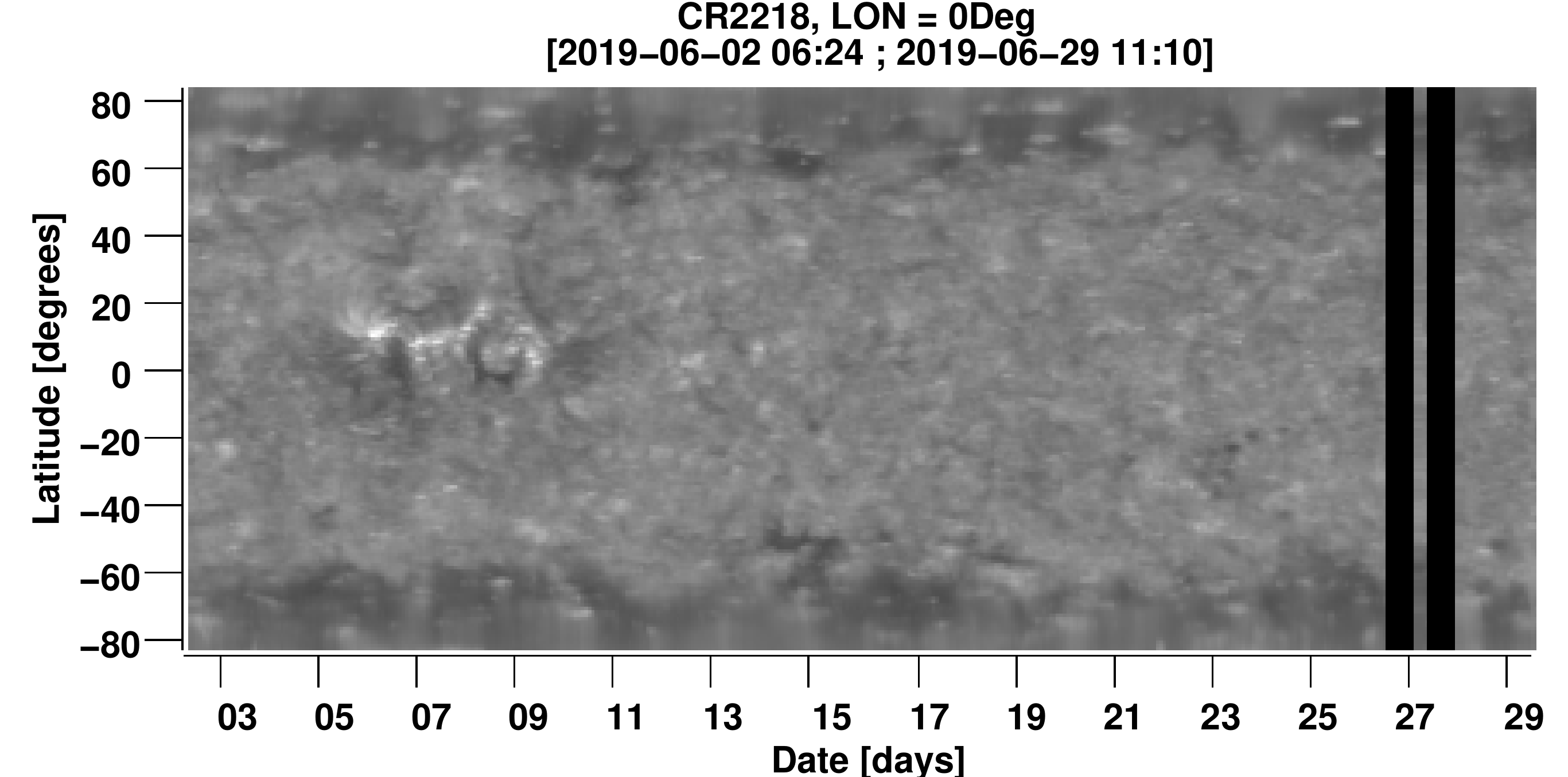}

\caption{SWAP synoptic maps for six individual CRs (2094: \textit{upper-left panel}, 2115: \textit{upper-right panel}, 2147: \textit{middle-left panel}, 2157: \textit{middle-right panel}, 2195: \textit{lower-left panel} and 2218: \textit{lower-right panel}). The \textit{black vertical stripes} indicate missing data. The horizontal axis represents the time in days from the \textit{first date mentioned at the top of the map} and the \textit{vertical axis} represents the Stonyhurst latitude in degrees. The \textit{vertical black stripe} for the \textit{middle-left} image represents missing data due to an off-pointing campaign on 26\,--\,27 February 2014.} \label{F-swapsynoptic}
\end{figure}

We characterize the evolution of the on-disk corona using SWAP synoptic maps, which reveal the evolution of activity at different latitudes with respect to time.
Figure \ref{F-swapsynoptic} shows six of these SWAP synoptic maps (or latitude--time maps) for six Carrington rotations (CR 2094: 27 February 2010 to 26 March 2010, CR 2115: 22 September 2011 to 19 October 2011, CR 2147: 11 February 2014 to 11 March 2014, CR 2157: 11 November 2014 to 8 December 2014, CR 2195: 23 September 2017 to 10 October 2017, CR 2218: 2 June 2019 to 29 June 2019). The vertical axis represents the Stonyhurst latitude in degrees and the horizontal axis the time in days. The way to build the synoptic maps is described in Appendix~\ref{S-synoptic}. Note that these maps are corrected for the heliographic latitude B$_\mathrm{0}$. This value is about -7$^\circ$ for the February\,--\,March timeframe (CRs 2094 and 2147), about +7$^\circ$ for September\,--\,October (CRs 2115 and 2195), about +2$^\circ$ for November\,--\,December (CR 2157) and about +1$^\circ$ for June (CR 2218).

From Figure~\ref{F-swapsynoptic} one can observe that northern hemispheric activity dominates for CR 2115 and the southern one dominates for CRs 2147 and 2157. In 2010 and 2019 only a few ARs are observed on the disk. The black vertical stripes in Figure~\ref{F-swapsynoptic} indicate observational gaps, for example due to off-pointing campaigns.

From the synoptic maps one can also follow the evolution of CHs. The coronal holes at the North Pole were present from February 2010 to October 2011, with some short intermittent periods. No CHs were observed between November 2011 and June 2015, with some short intermittent periods also. They started to develop again in July 2015 and remained visible until June 2019 (end of our dataset). At the south pole the CHs were present from February 2010 to May 2012, with some intermittent periods. No CHs were observed between June 2012 and May 2014. They started to develop again in June 2014 and remained visible until June 2019.

The development and disappearance of CHs are measured manually. To mitigate any observer bias, four independent sets of measurements were made manually (referred to as observers 1,2,3 and 4), and the fifth was made automatically. A summary of these five case-studies are presented in Table~\ref{T-chs}. The first three measurements were made by three separate observers using SWAP synoptic maps. The fourth observer used AIA+EUVI synoptic maps, while the fifth set of observations were made using an automated detection algorithm.

Due to the spread in measurements made by the first four observers, the manual measurements were cross-checked against automated coronal hole detection algorithms, for magnetic field characteristics, such as polarity imbalance. For this we used the CHIMERA \citep{Garton2018} and CHARM  \citep{Krista2009} databases, which are available on the solar monitor: \url{www.solarmonitor.org}. CHIMERA covers the period from 2 September 2010 to 30 June 2019, while CHARM is used for the seven months from 3 February to 1 September 2010.

CHIMERA was developed for use with SDO/AIA \citep{Lemen2012} observations at wavelengths of 17.1\,nm, 19.3\,nm, and 21.1\,nm and with magnetograms from the \textit{Helioseismic and Magnetic Imager} \citep[HMI, ][]{Scherrer2012} onboard SDO. The algorithm is based on a multi-thermal intensity segmentation technique. As CHs have a relatively low temperature and density, they are observed as dark features in EUV images. A higher contrast relative to the ambient solar corona is observed in the 17.1 nm passband compared with the other passbands. The Magnetograms are used to distinguish CHs from other, smaller non-CH regions depending on if they exhibit a unipolar magnetic structure. Most segmentation algorithms use observations of the solar disk, however CHIMERA is also capable of detecting off-limb, high altitude components of coronal holes. This is valuable for detecting back-sided coronal holes.

CHARM identifies coronal holes using a histogram-based intensity thresholding technique, using EUV and X-ray observations from the imagers on STEREO \citep{Kaiser2008}, SOHO \citep{Domingo1995} and Hinode \citep{Kosugi2007}. The histograms are used to determine the threshold for low intensity regions, which are then classified as coronal holes or filaments using magnetograms from the SOHO/MDI instrument.

Based on the CHIMERA data, from the period October 2010 to December 2015, coronal holes above 70 degrees or with extensions above 70 degrees in latitude were tracked. It was observed that the northern polar hole dissipated in December 2011 and reappeared in September 2014. In contrast, the southern polar CH disappeared around February 2013, and reappeared in February 2014.

As described in \citep{Cranmer2009}, the development of such polar CHs starts at lower latitudes (40-50 degrees), at earlier times, before migrating toward the polar regions. The CHIMERA database shows that the northern CH appeared at lower latitudes in November 2011, whereas the southern CH appeared around January 2012. However, the evolution to the poles is quite different. In the southern hemisphere, smaller CHs are observed to move quickly to the polar region (above 70 degrees), growing in size and creating a persistent polar coronal hole by February 2014. However, in the northern hemisphere, the CHs developed into much larger sizes at lower latitudes (about twice the size of the southern "polar" CHs), but moved much slower towards the pole, resulting in a long-lasting polar CH (above 70 degrees) from September 2014. It is worth noting that during the development of the polar coronal holes, at lower latitudes, CHs of opposing polarity can be seen (e.g. October-November 2014).

Note that CHIMERA was specifically designed to retrieve coronal holes and measure their characteristics (size, magnetic flux, etc.), whereas SWAP synoptic maps show only the changes in the intensity around the 17.4\,nm wavelength. As a consequence, the detection of CHs in SWAP data is based on a visual detection of dark regions at the poles. This may account for the disparity between the manual measurements presented in Table~\ref{T-chs}. It also explains why polar CHs are visible longer and are seen to start developing earlier by CHIMERA than in the SWAP images.

In this study we take the dates from the first observer, which were described above.\\

\begin{table}
	\caption{Coronal hole evolution in SC24. The time is in format yyyy.mm. *Movie available at \url{www.stce.be/news/423/welcome.html}} \label{T-chs}
	\begin{tabular}{cccc}
		\hline
		  & N   & S   & Observer/Source   \\
		\hline
		CHs gone &  2011.11 & 2012.06 & Observer 1 / SWAP synoptic maps \\
		CHs development & 2015.07 & 2014.06 & Observer 1 / SWAP synoptic maps \\
		\hline
		CHs gone & 2011.11  & 2012.04 & Observer 2 / SWAP synoptic maps \\
	    CHs development & 2015.10 & 2015.10 & Observer 2 / SWAP synoptic maps \\
		\hline
		CHs gone & 2011.11  & 2012.09 & Observer 3 / SWAP synoptic maps \\
		CHs development & 2015.05 & 2015.03 & Observer 3 / SWAP synoptic maps \\
		\hline
		CHs gone & 2011.05  & 2012.08 & Observer 4 / AIA+EUVI synoptic maps*   \\
		CHs development & 2015.05 & 2014.05 & Observer 4 / AIA+EUVI synoptic maps*  \\
		\hline
		CHs gone & 2011.12  & 2013.02 & CHIMERA / AIA+HMI   \\
		CHs development & 2014.09 & 2014.02 & CHIMERA / AIA+HMI  \\
		\hline
		
	\end{tabular}
\end{table}

\textit{SWAP East--West Synoptic Maps}\\

Similar to the SWAP synoptic maps, SWAP East--West (EW) synoptic maps were constructed and examples can be seen in Figure~\ref{F-ewswapsynopticlat} (see Appendix~\ref{S-ondisk} for explanations).

Figure~\ref{F-ewswapsynopticlat} shows the SWAP EW synoptic maps (or time--longitude maps) for CRs 2094, 2115, 2147, 2157, 2195, and 2218, at the solar Equator. The horizontal axis represents the Stonyhurst longitude (in degrees) and the vertical axis represents the time in days from the first date mentioned at the top of the map. 
As the maps span the longitude from East to West limb, any long-lasting features on the solar disk (\textit{e.g.}, bright ARs or dark filaments/CHs) can be seen as bright or dark stripes, respectively, in these synoptic maps.

\begin{figure}
\centering
\includegraphics[height=0.55\textwidth]{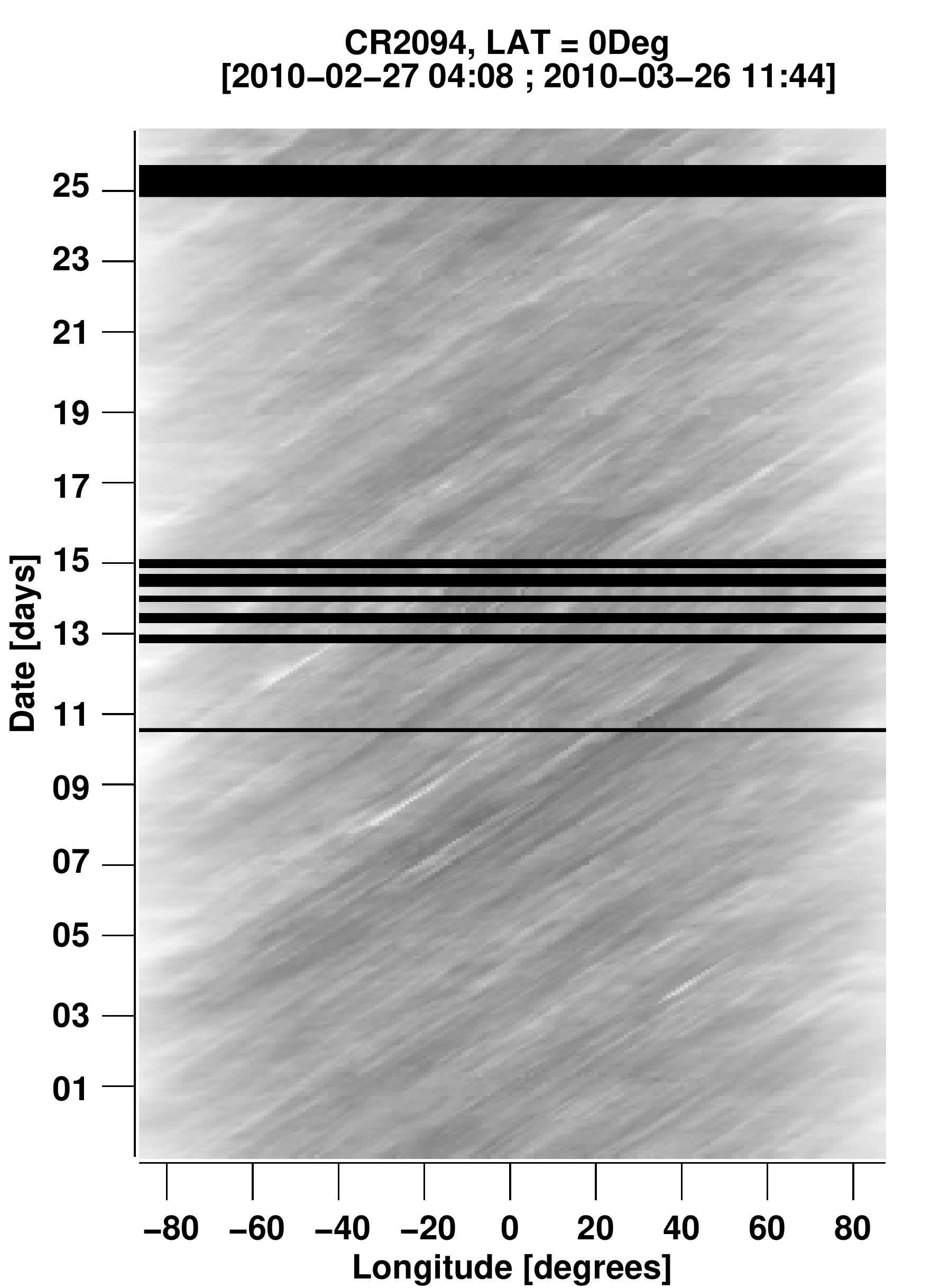}
\includegraphics[height=0.55\textwidth]{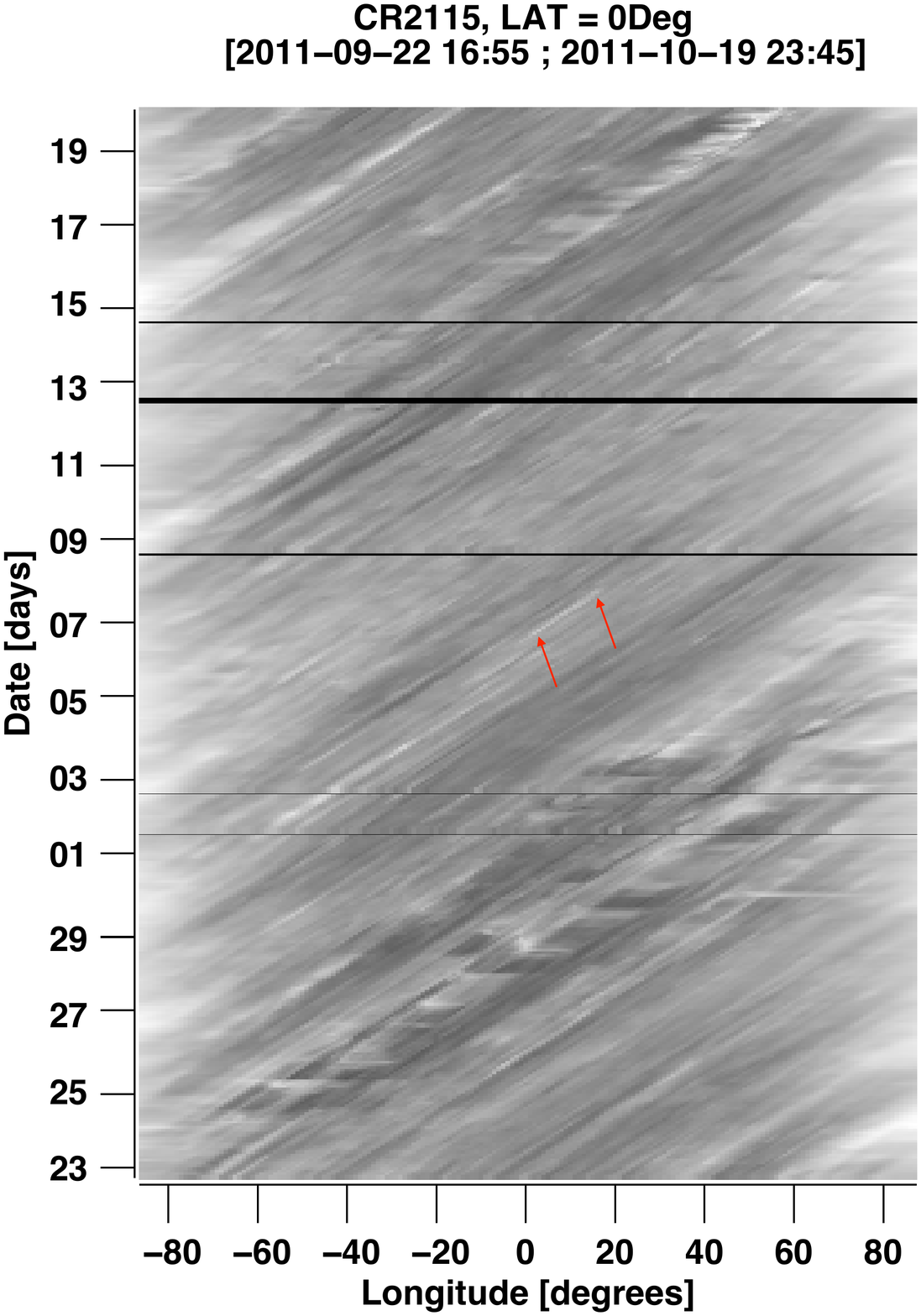}
\includegraphics[height=0.55\textwidth]{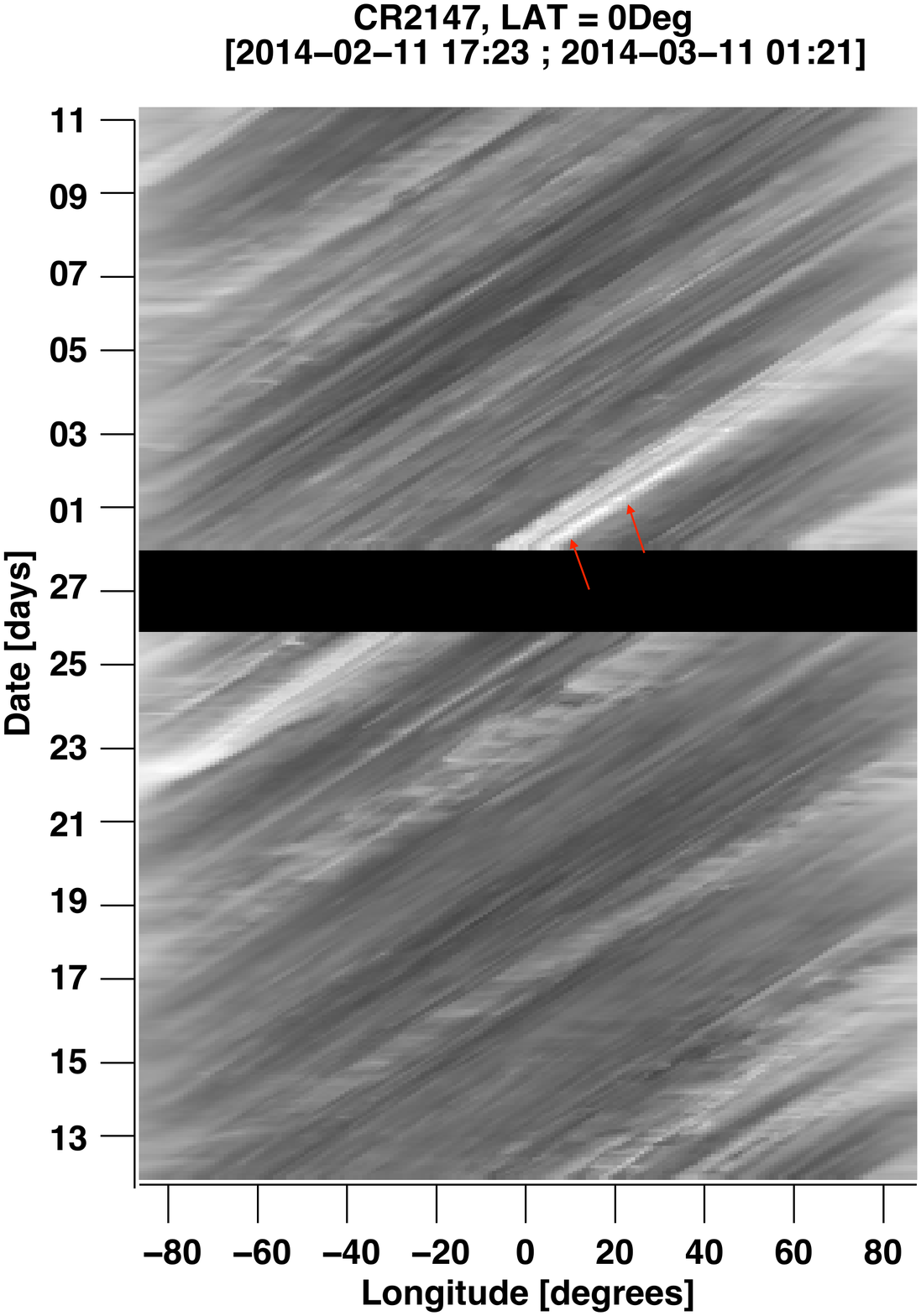}
\includegraphics[height=0.55\textwidth]{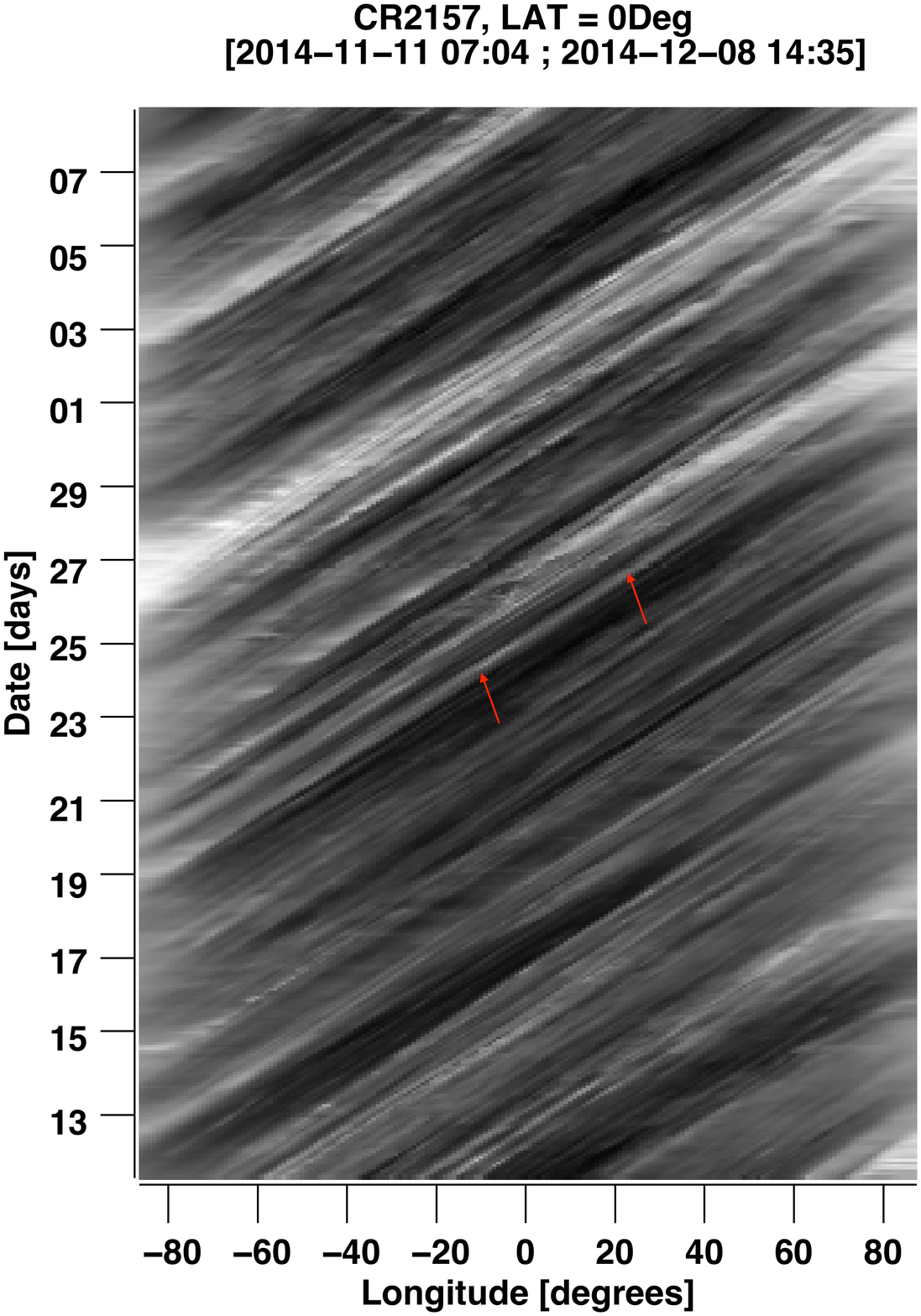}
\includegraphics[height=0.55\textwidth]{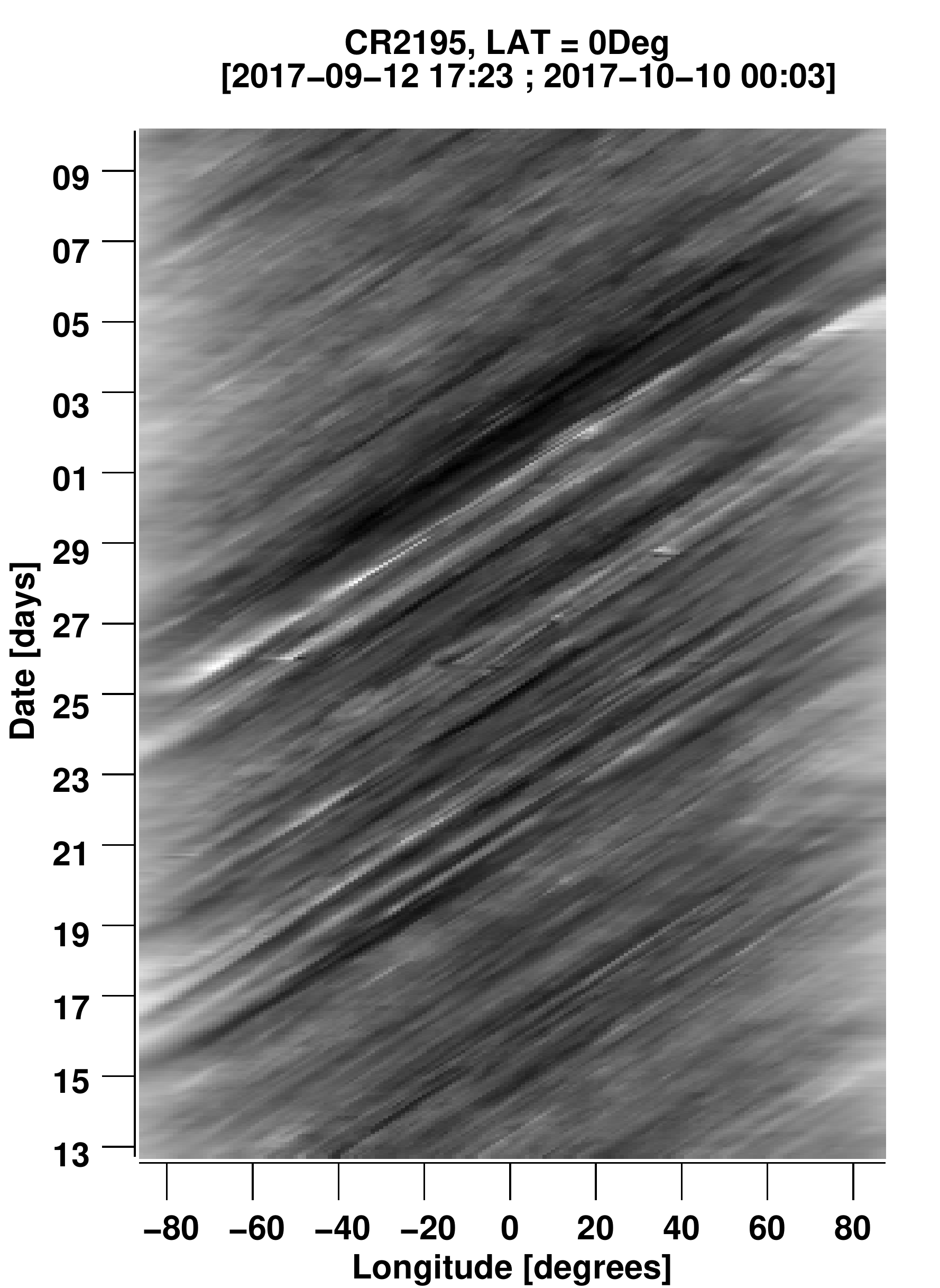}
\includegraphics[height=0.55\textwidth]{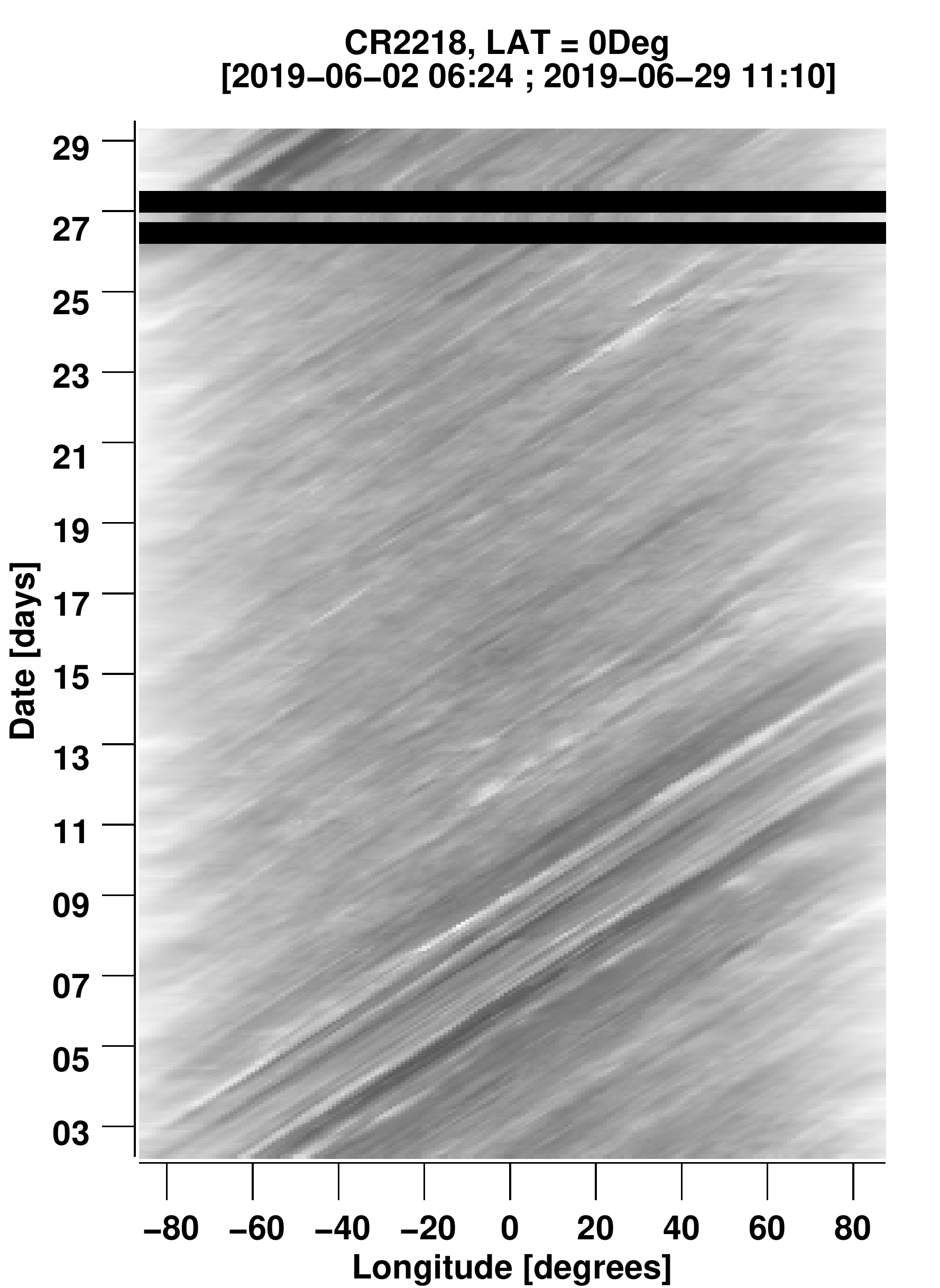}

\caption{SWAP EW synoptic maps for the equatorial region of six CRs: 2094, 2115, 2147, 2157, 2195 and 2218. The \textit{horizontal axis} represents the Stonyhurst longitude [degrees] and the \textit{vertical axis} represents the time in days from the \textit{first date mentioned at the top of the map}. The \textit{red arrows} point to three tracked stripes. The maximum estimated errors are around 2.4\,degree\,day$^{-1}$ - see text for details.}
\label{F-ewswapsynopticlat}
\end{figure}

We see that some bright stripes span the full longitude range, from -90$^{\circ}$ to 90$^{\circ}$, while some appear and/or disappear at different longitudes, suggesting the birth and/or fading of the ARs at those locations.
By following the bright stripes we can estimate the dynamics of the solar features in terms of average rotation at different latitudes. The measurements were done by selecting manually two points along a given stripe (\textit{e.g.}, see the red arrows in Figure~\ref{F-ewswapsynopticlat}) and by calculating the rotation speed from these two points, as the ratio between \textit{(l2-l1)} and \textit{(t2-t1)}. Here \textit{l2} and \textit{l1} are the heliographic longitudes of the second and first point and \textit{t2} and \textit{t1} are the times of the second and first point respectively. The units are in [\textit{degree\,day$^{-1}$}]. The average rotation rates were estimated from points having longitudes between -30$^{\circ}$ and +30$^{\circ}$ to avoid limb effects.

In order to estimate the errors we have repeated the measurements five times, for a series of different types of stripes (i.e. short and thin stripes or long and well-defined stripes etc.). We got errors up to 2.4\,degree\,day$^{-1}$ for the very small, thin stripes and errors of approximately 0.1\,degree\,day$^{-1}$. for the well defined, bright stripes. Even for the wide 5 degrees stripes, the errors were quite small, in the order of 0.13\,degree\,day$^{-1}$. In general we tried to avoid measurements when big data gaps were present (like CR2147 in Figure~\ref{F-ewswapsynopticlat}), i.e. we selected stripes which were not affected by these data gaps. We calculated the errors for stripes where data gaps were smaller (see e.g. CR 2115 in Figure~\ref{F-ewswapsynopticlat}) and we got values of around 1.2\,degree\,day$^{-1}$.

Table~\ref{T-rotation} gives a summary of these measurements for different EW synoptic maps at different latitudes. The average rotation speed of solar features observed between latitudes of -40$^{\circ}$ and +40$^{\circ}$ is around 14\,degree\,day$^{-1}$. Note that the bright stripes may indicate either ARs (loops) or bright points or large-scale bright coronal features that can have a large latitudinal extent. The EUV images display the line-of-sight integrated emission of the solar corona, and as a consequence, a superimposition of different features may be observed in the same stripe.

\begin{table}
	\caption{Rotation rates [degree\,day$^{-1}$] of bright features at five different latitudes: S40, S20, 0, N20 and N40 and their average values. The values with stars indicate a very small stripe from which the rotation speed was calculated. The maximum estimated errors are around 2.4\,degree\,day$^{-1}$ - see text for details. The format for CR starting date is yyyy.mm} \label{T-rotation}
	\begin{tabular}{cccccccc}
		\hline
		CRs  & CR starting date & S40   & S20   & 0     & N20   & N40 & Mean \\
		\hline
		2094 & 2010.02 & 12.87 & 14.51 & 12.34* & 13.15 & 13.51*  & 13.27 \\
		2115 & 2011.09 & 13.69* & 13.32 & 14.70* & 14.13 & 12.62* & 13.69 \\  
		2147 & 2014.02 & 13.88* & 13.55 & 15.00 & 13.45 & 15.36*  & 14.24 \\ 
		2157 & 2014.11 & 13.77 & 13.57 & 14.90 & 13.52 & 13.25    & 13.80 \\
		2195 & 2017.09 & 15.02* & 14.47 & 13.44 & 13.88 & 14.49*  & 14.26 \\
		2218 & 2019.06 & 12.64* & 14.20* & 15.13 & 15.33 & 15.35* & 14.53 \\
		Mean & - & 13.64 & 13.93 & 14.25 & 13.91 & 14.09    & \textbf{13.96} \\
		
	\end{tabular}
\end{table}

\begin{table}
	\caption{Rotation rates [degree\,day$^{-1}$] of bright features at three different latitudes: S15, 0, N15 for CRs in June, for each year from 2010 to 2019. The values with stars indicate a very small stripe from which the rotation speed was calculated. The CR starting date is in the format yyyy.mm} \label{T-rotationyear}
	\begin{tabular}{cccccc}
		\hline
		CRs  & CR starting date & S15    & 0      & N15     & Mean      \\
		\hline
		2098 & 2010.06 & 13.65  & 15.51* &  14.15  & 14.43      \\
		2111 & 2011.06 & 14.39  & 14.70* &  14.89  & 14.66      \\  
		2124 & 2012.06 & 14.74  & 15.80  &  14.35  & 14.96      \\ 
		2138 & 2013.06 & 14.22  & 14.42  &  13.97  & 14.20      \\
		2151 & 2014.06 & 14.22  & 14.33  &  14.28  & 14.27      \\
		2164 & 2015.06 & 14.98  & 15.72  &  14.60  & 15.10      \\
		2178 & 2016.06 & 14.34  & 16.16* &  13.80* & 14.76      \\
		2191 & 2017.06 & 15.07* & 16.15* &  15.16  & 15.46      \\
		2205 & 2018.06 & 15.36* & 14.81  &  14.94  & 15.03      \\
		2218 & 2019.06 & 13.77* & 15.12  &  14.70  & 14.53      \\
		Mean & - & 14.47  & 15.27  &  14.48  & \textbf{14.74} \\
		
	\end{tabular}
\end{table}

To estimate the evolution of the dynamics of the solar features throughout SC24 we take for each year a CR in June (when B$_\mathrm{0}$ is around 0$^{\circ}$) and calculate the average rotation rate at three latitudes: +15$^{\circ}$, 0$^{\circ}$, and -15$^{\circ}$ separately. The choice of these latitudes is based on a quick check of a number of 1677 NOAA ARs that appeared between January 2010 and June 2018. The northern groups peak at a latitude of +13$^{\circ}$, and the southern groups peak at -17$^{\circ}$.

The estimated values of the rotation rates for the three latitudes are displayed in Table~\ref{T-rotationyear}, and they are around 15\,degree\,day$^{-1}$ throughout the whole SC24. Note that these values indicate the rotation of different line-of-sight integrated features across the solar disk (\textit{e.g.} ARs, loops, streamers, fans, bright points), as mentioned above. Some of the bright stripes in the EW synoptic maps may also be due to spotless regions. 

On top of this, the errors of the estimated speeds are quite big due to pointing errors in selecting the measurement points, caused by different types of stripes: short, thin, large, faint, bright etc. Missing data adds to this errors as well. The uncertainties can reach values up to 2.4\,degree\,day$^{-1}$.
The deduced mean rotation rates for features at latitudes of $\pm$15$^{\circ}$ are around 14.5\,degree\,day$^{-1}$, and they compare very well with the current set of accepted values for the average photospheric rotation rates (see, \textit{e.g.}, \citealt{Snodgrass1990}), respectively 14.475\,degree\,day$^{-1}$ (South) \textit{vs.} 14.545\,degree\,day$^{-1}$ (North).

\begin{figure}
\includegraphics[height=0.45\textwidth]{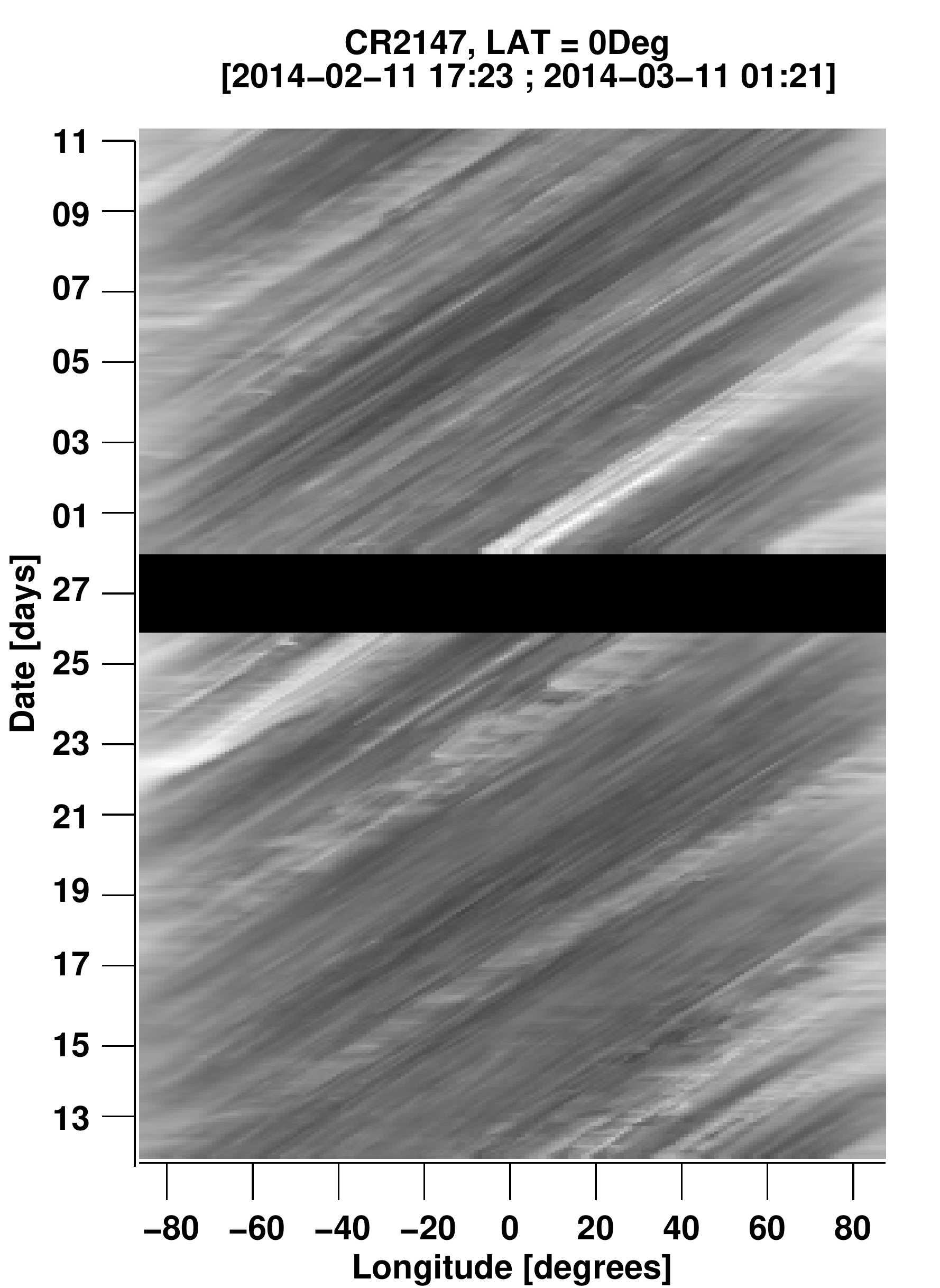}
\includegraphics[height=0.45\textwidth]{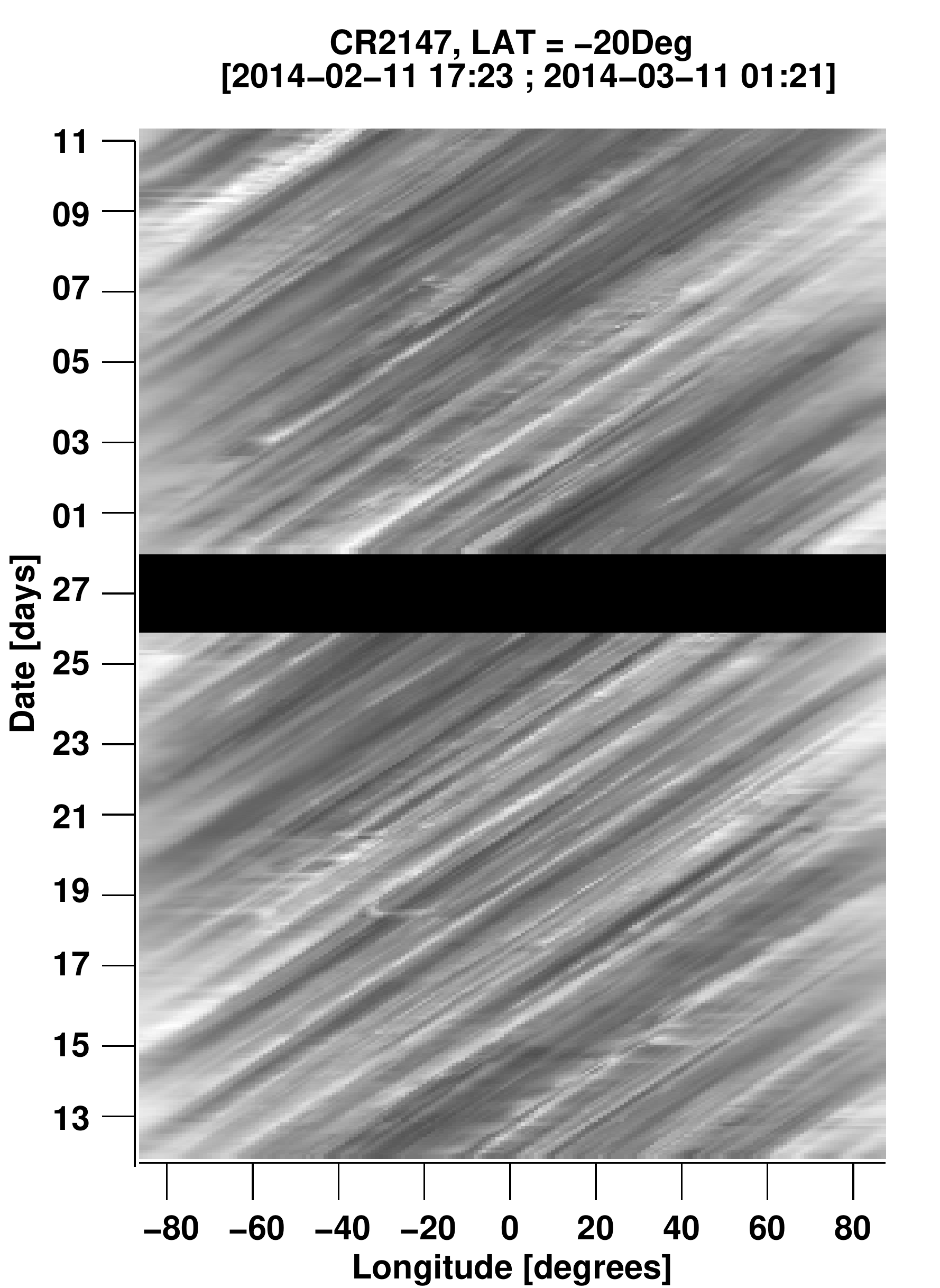}
\includegraphics[height=0.45\textwidth]{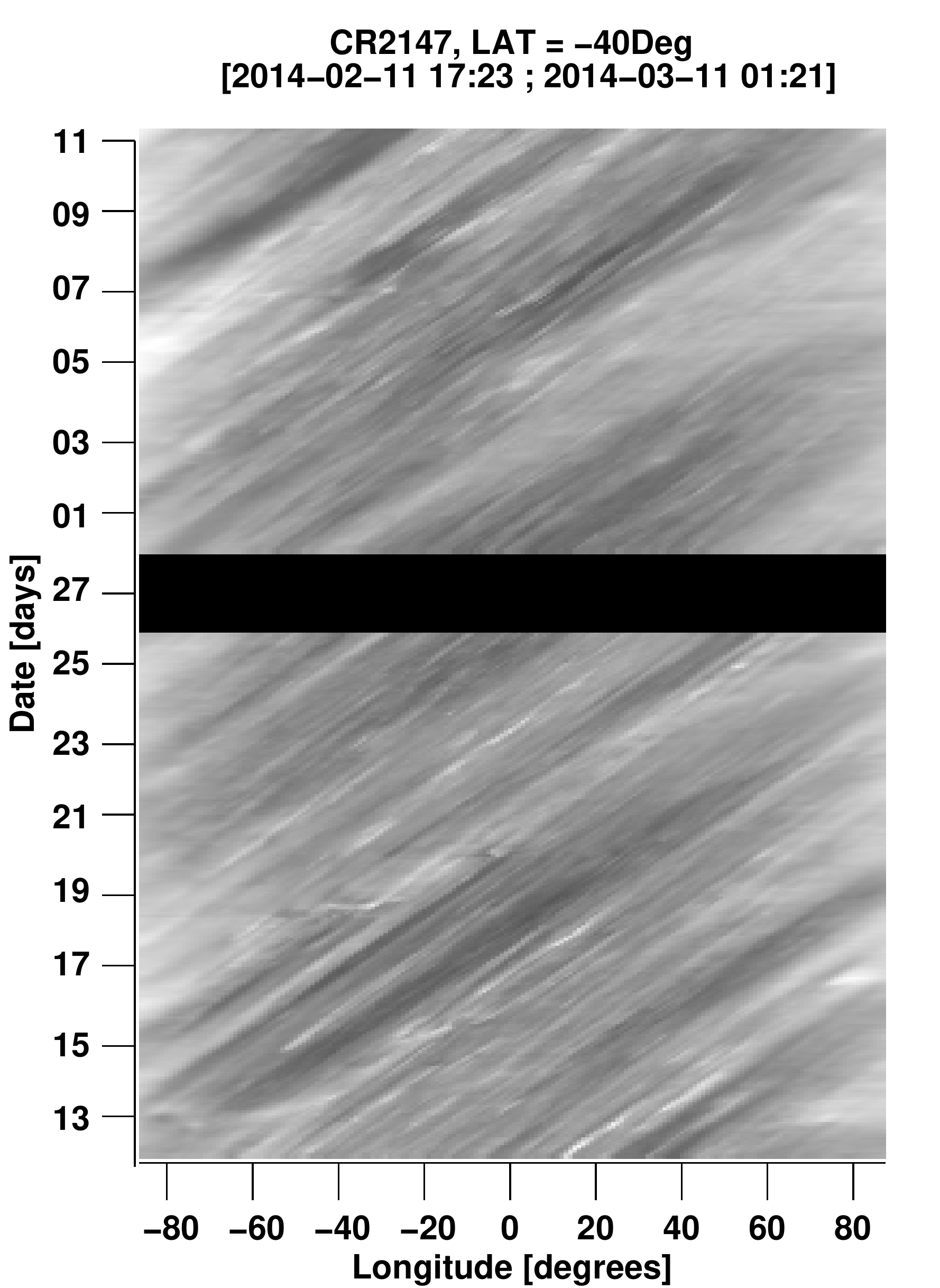}
\includegraphics[height=0.45\textwidth]{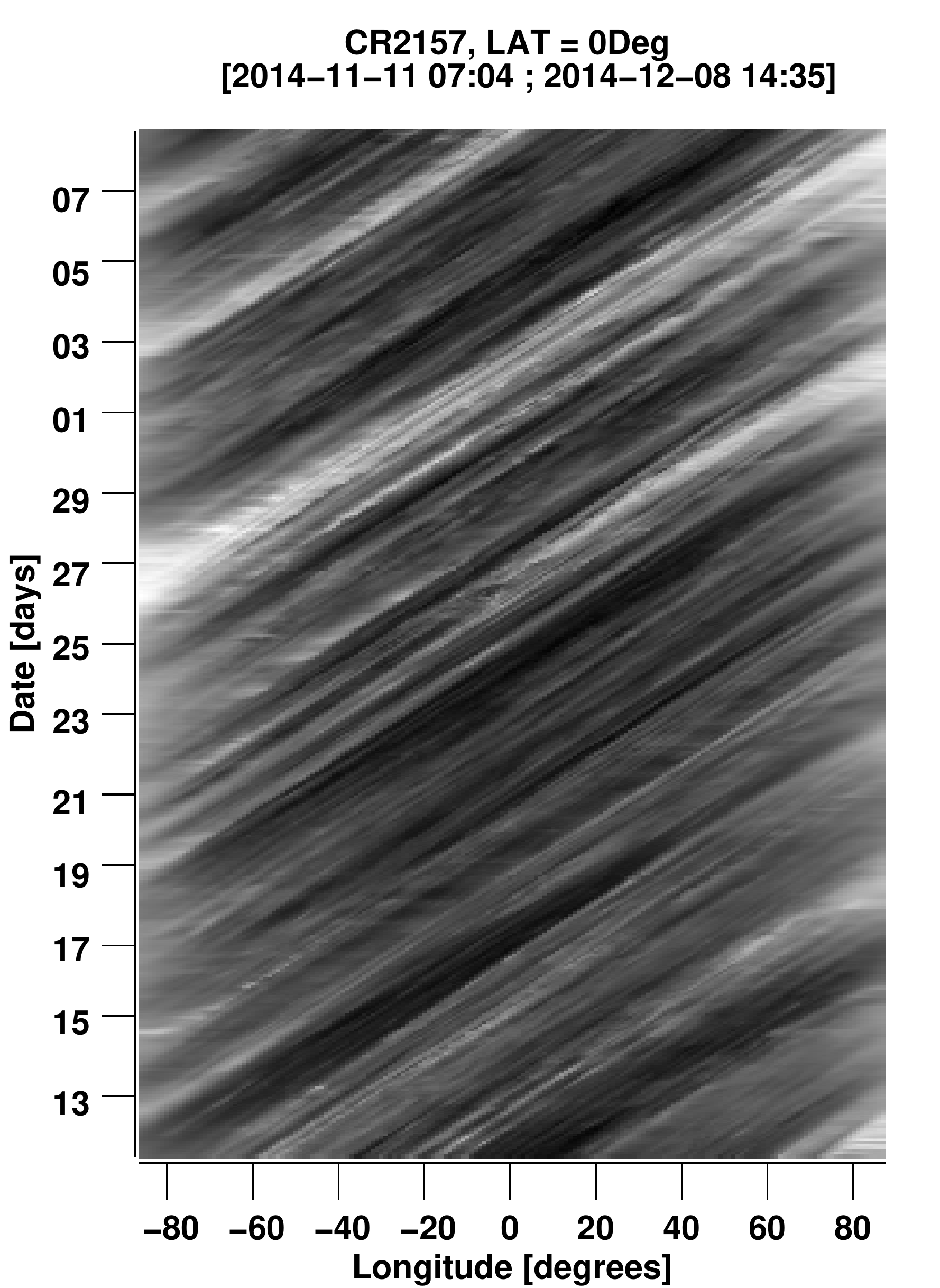}
\includegraphics[height=0.45\textwidth]{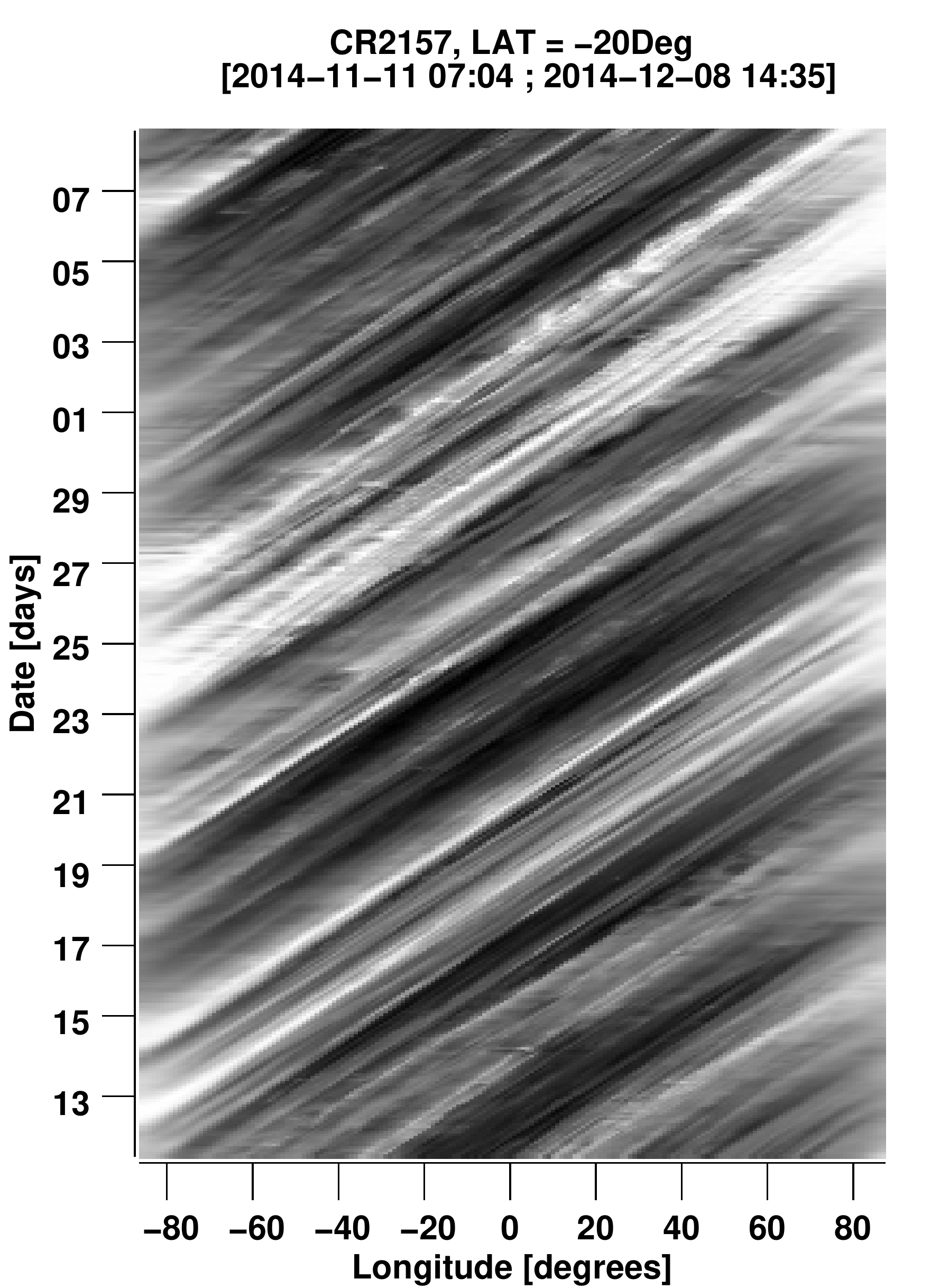}
\includegraphics[height=0.45\textwidth]{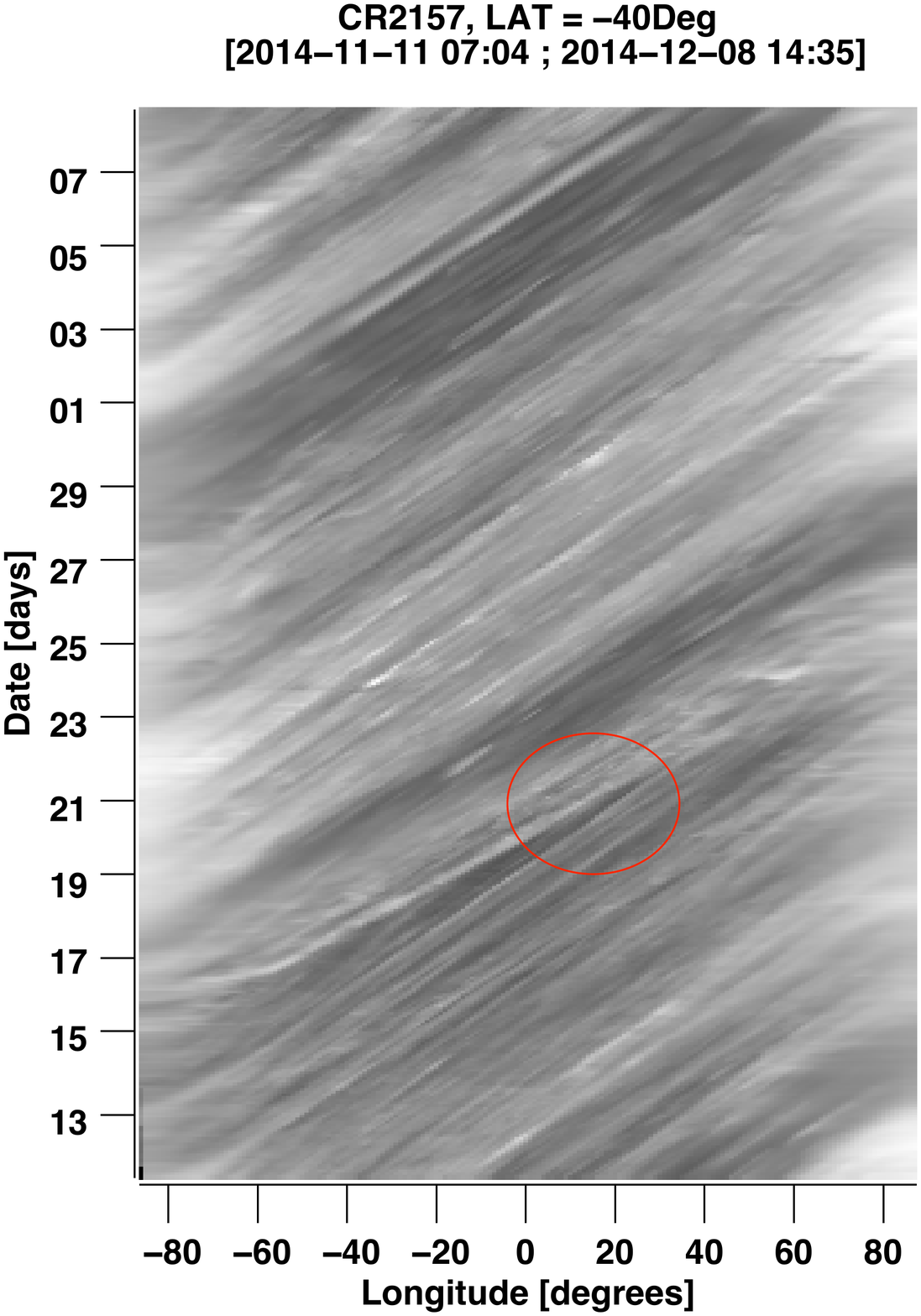}

\caption{SWAP EW synoptic maps for CR2147 (\textit{upper panels}) and CR2157 (\textit{lower panels}) at Equator (\textit{left panels}), 20$^{\circ}$ South (\textit{middle panels}) and 40$^{\circ}$ South (\textit{right panels}). The \textit{horizontal axis} represents the Stonyhurst longitude [degree] and the \textit{vertical axis} represents the time in days from the first date mentioned at the top of the map. The \textit{red circle} shows the converging stripes - see text for details.} \label{F-ewswapsynoptic2157}
\end{figure}

Figure~\ref{F-ewswapsynoptic2157} shows SWAP EW synoptic maps for CR2147 and for CR2157 at the Equator (upper and lower-left panels respectively), 20$^{\circ}$ south (upper and lower-middle panels) and 40$^{\circ}$ south (upper and lower-right panels). The horizontal axis represents the Stonyhurst longitude (in degree) and the vertical axis represents the time in days from the first date mentioned at the top of the image. The different inclination of the stripes indicate different rotation speeds for the solar features at those latitudes. Note that in each map the bright stripes are mostly parallel (an indication of a constant rotation speed at that latitude), but sometimes one can see also converging stripes (\textit{e.g.} lower-right panel of Figure~\ref{F-ewswapsynoptic2157} enclosed by the red circle). This indicates a bright feature rotating with a different speed at that latitude (40$^{\circ}$ South) or a large-scale feature rooted at a different latitude and visible also at 40$^{\circ}$ South. 

\subsubsection{The Evolution of the Off-Limb Solar Corona}

To characterize the evolution of the off-limb region we create off-limb synoptic maps -- see explanations in Appendix~\ref{S-offlimb}.
Figure~\ref{F-offlimbimg} shows the off-limb corona at the South Pole for CR 2157 (11 November 2014 to 8 December 2014) at three different distances from the Sun's center: 1.1, 1.3, and 1.6\,R$_{\odot}$. The time frame coincides with the pseudostreamer observed at the south pole and discussed by \citet{Guennou2016}. In this figure one can see the bright inclined stripe from the lower-left side of the figure to the middle-right side of the figure, equivalent to a front-disk rotating large-scale feature, and the inclined bright stripe from the middle-right to the upper-left corner of the figure, equivalent to a behind-disk rotating large-scale feature.

\begin{figure}

\includegraphics[width=0.32\textwidth]{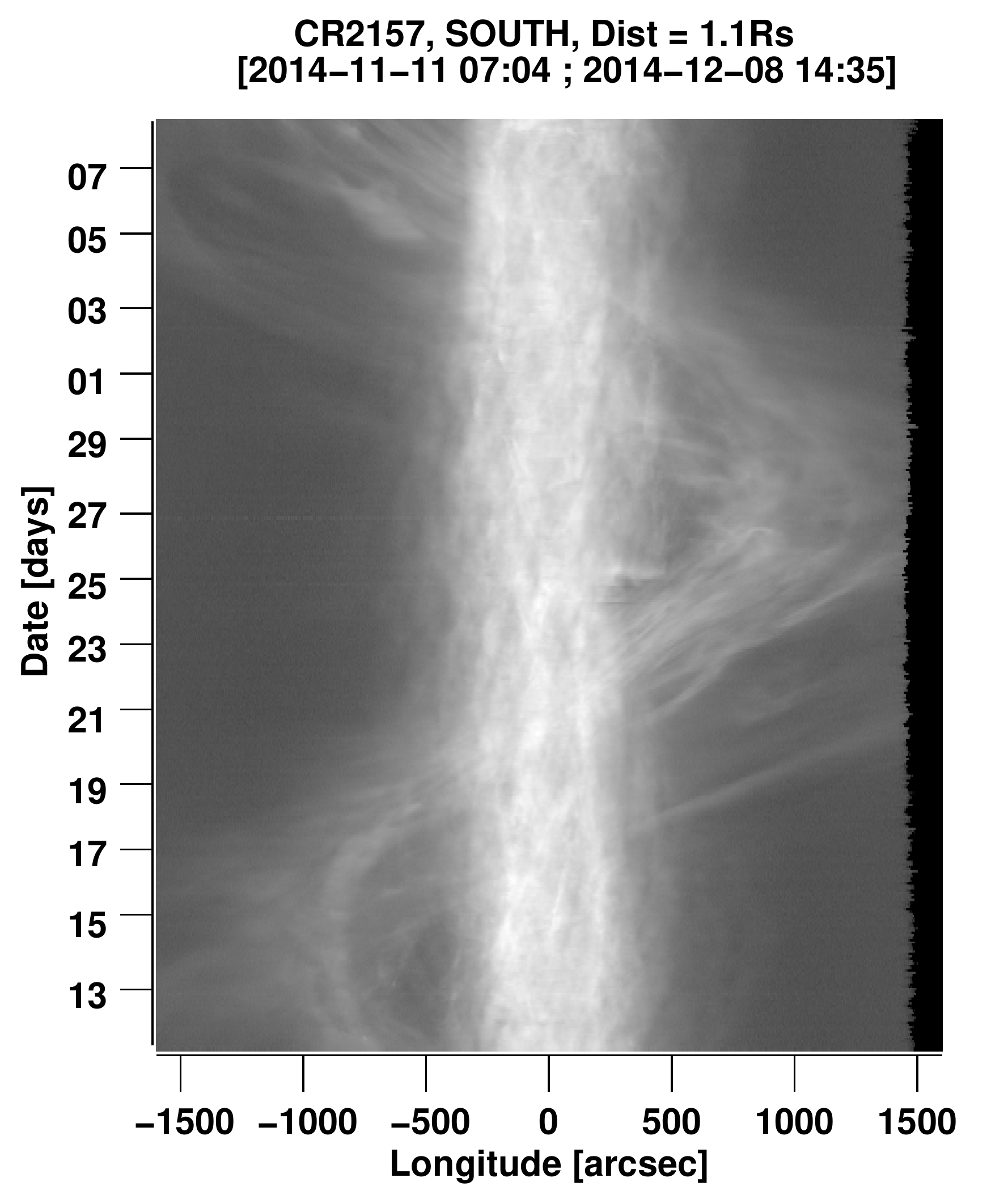}
\includegraphics[width=0.32\textwidth]{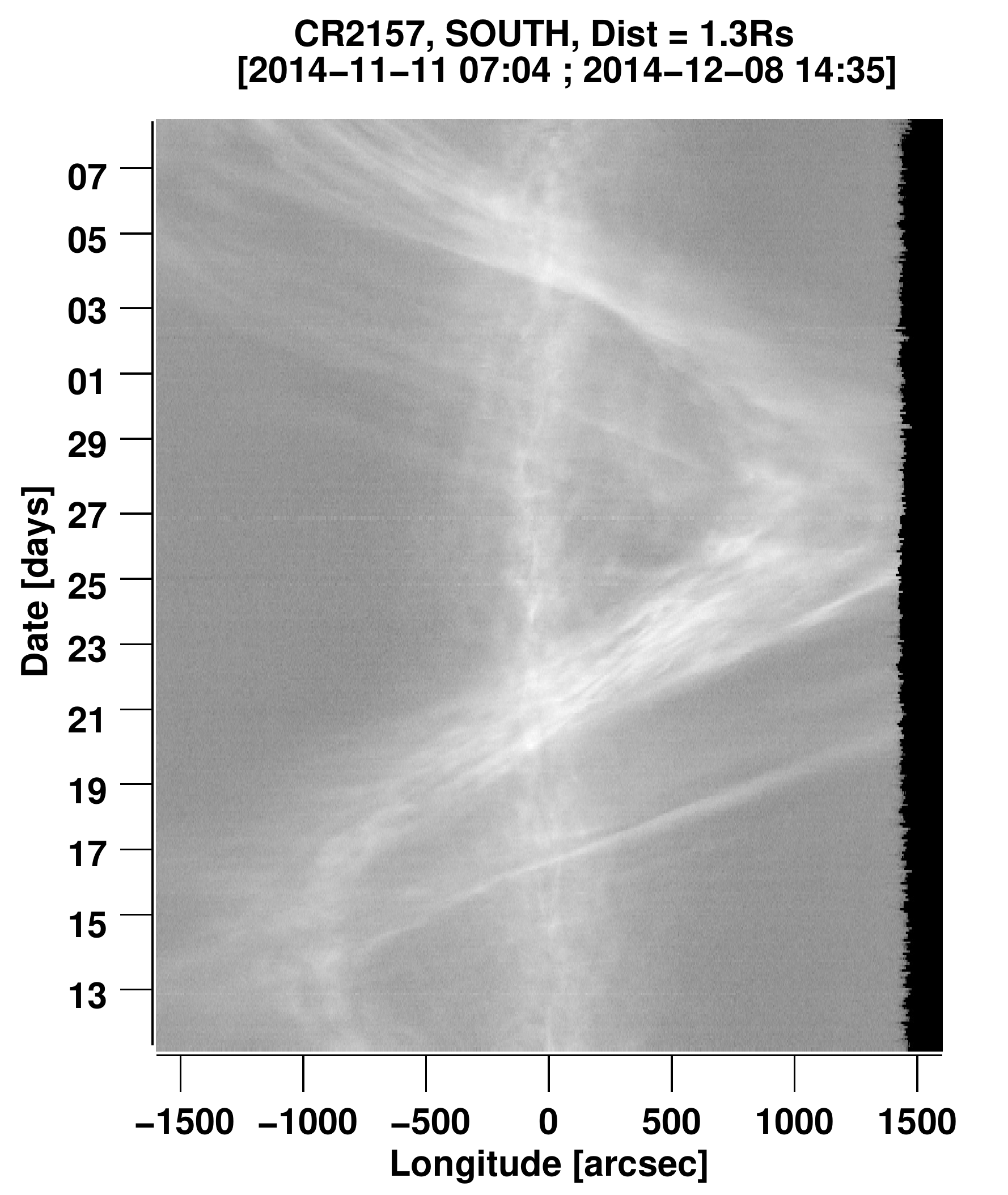}
\includegraphics[width=0.32\textwidth]{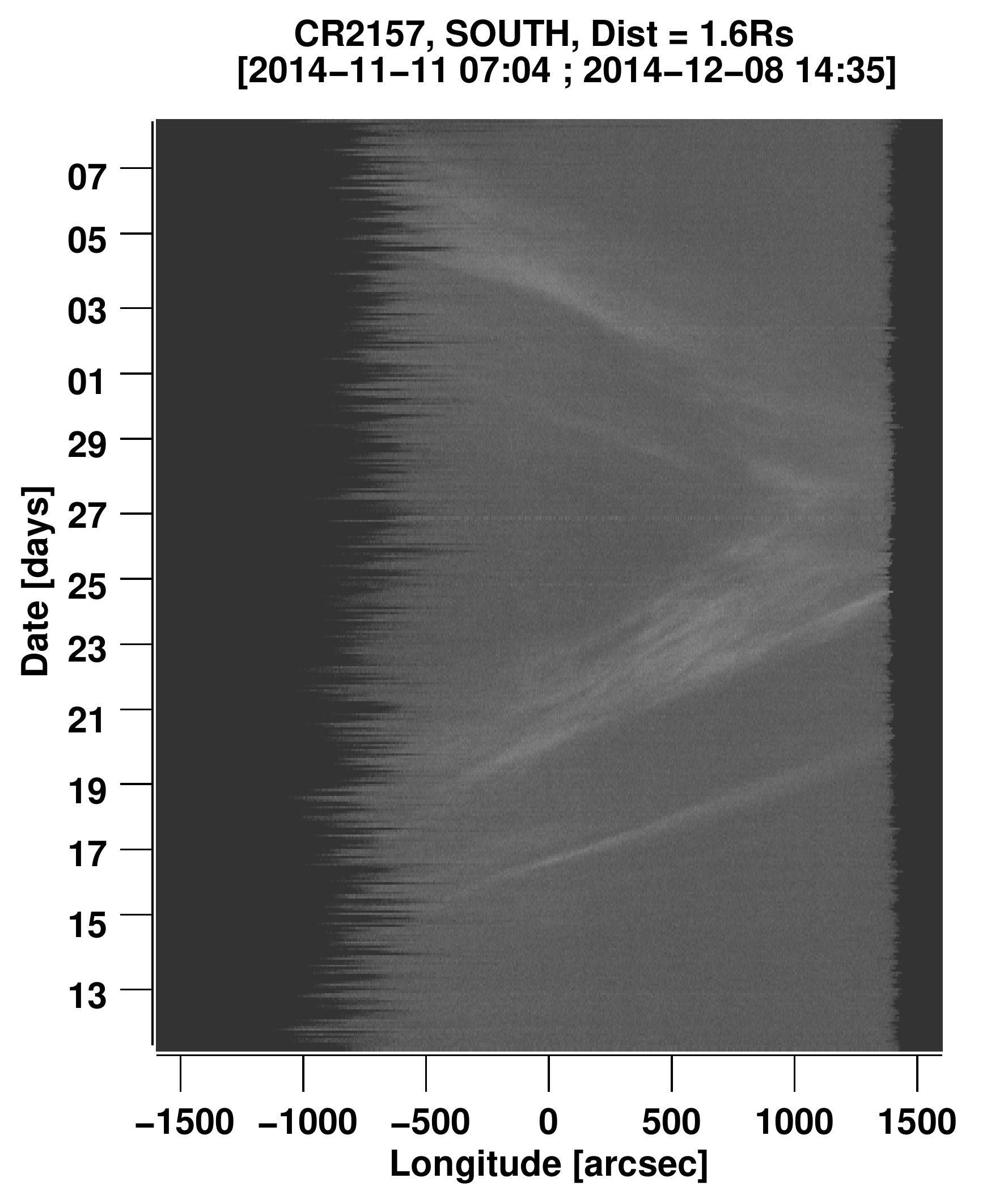}

\caption{SWAP EW off-limb synoptic map (or off-limb time--longitude maps) for CR2157 at South Pole for the central distance of 1.1, 1.3, and 1.6\,R$_{\odot}$ from the Sun center. The two sinusoidal patterns indicate the complexity of the large-scale structure observed at the pole on this period. The \textit{horizontal axis} represents the helioprojective longitude in arcseconds and the \textit{vertical axis} the time in days from the first date mentioned at the top of the map. The distance mentioned in the title of the figure is the shortest distance from the Sun center to the horizontal stripe used to build the synoptic map. The \textit{black-pixel regions} are due to no-signal in SWAP data, resulting from rotating the original images to bring solar North up.} \label{F-offlimbimg}
\end{figure}

\begin{figure}
	
	\includegraphics[width=0.48\textwidth]{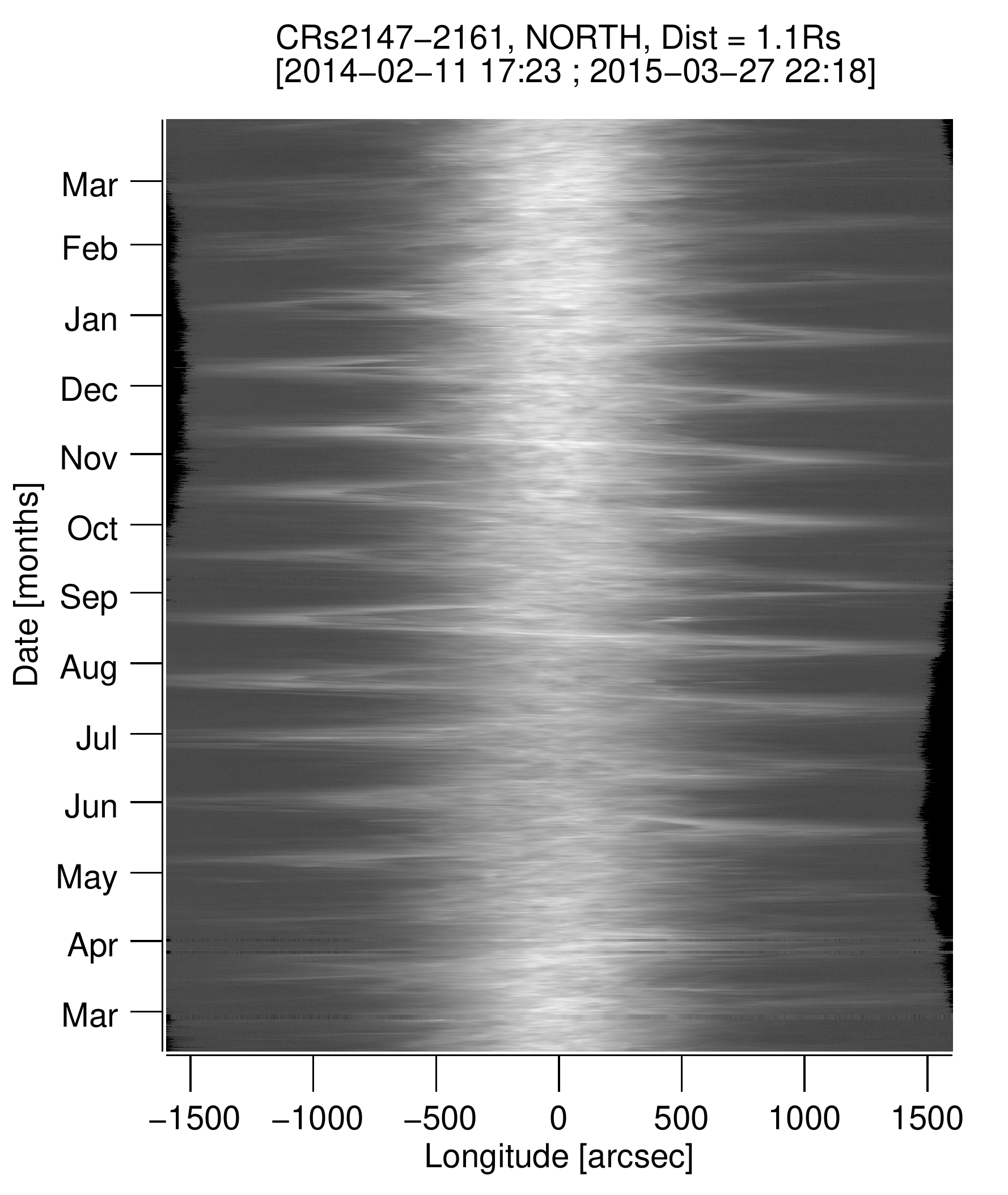}
	\includegraphics[width=0.48\textwidth]{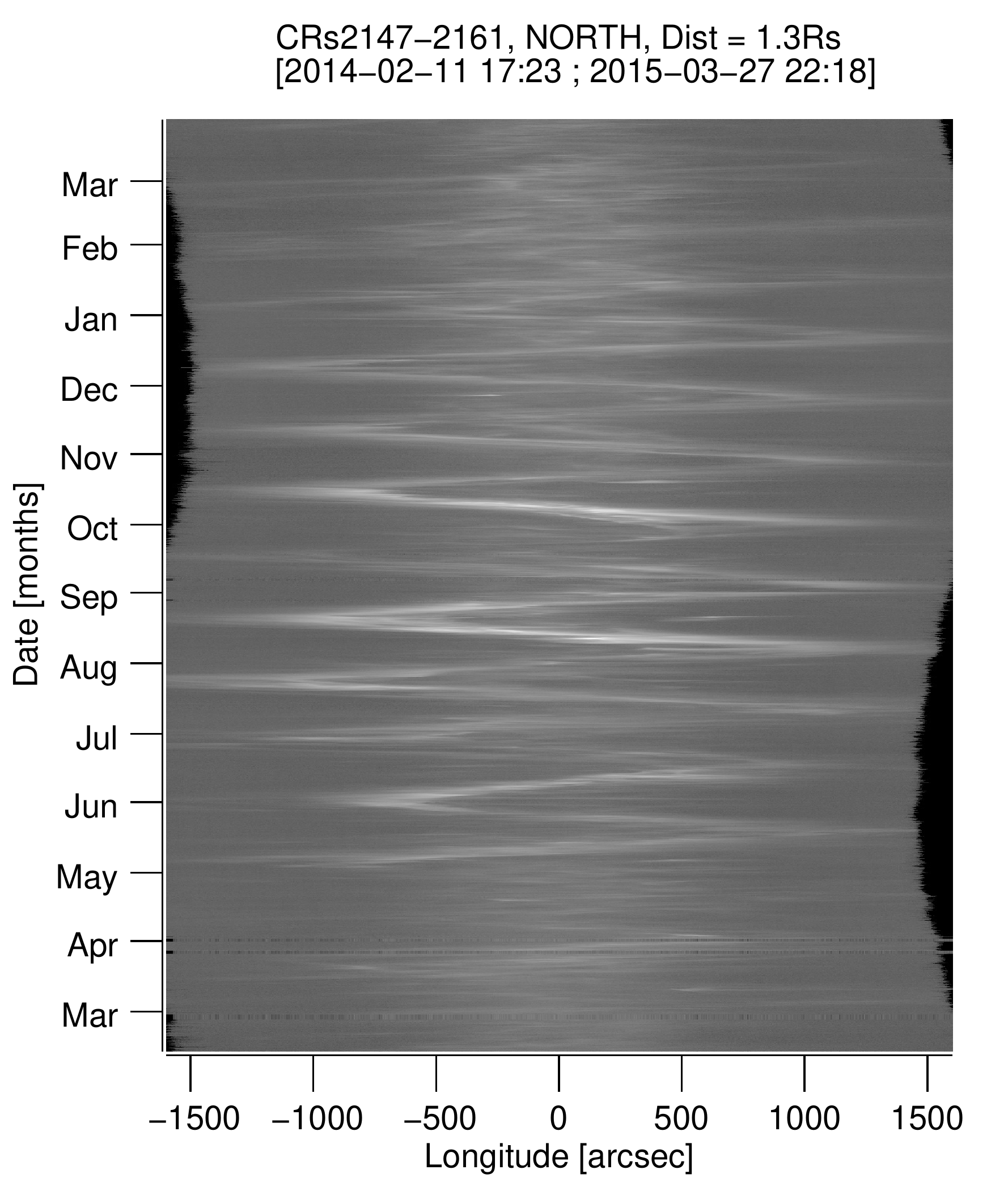}
	
	\caption{SWAP EW off-limb synoptic map (or off-limb time-longitude maps) for CRs 2147 to 2161 at North Pole for the central distance of 1.1 and 1.3 R$_{\odot}$ from the Sun center. The horizontal axis represents the helioprojective longitude in arcseconds and the vertical axis the time in months from the first date mentioned at the top of the map. The distance mentioned in the title of the figure is the shortest distance from the Sun center to the horizontal stripe used to build the synoptic map. The black pixel regions are due to no-signal in SWAP data, resulting from rotating the original images to bring solar North up.} \label{F-offlimbfans}
\end{figure}

Figure~\ref{F-offlimbfans} shows the off-limb corona at the North Pole for 15 CRs, from 2147 to 2161 (11 February 2014 to 27 March 2015) at two different distances from the Sun's center: 1.1 and 1.3\,R$_{\odot}$. We see that a large-scale feature, a coronal fan (see also \citealt{Talpeanu2016}), persists for more than 11 CRs. In the period when the fan was observed, the North Pole, off-limb, was dominated by bright ''spicules'', and many ARs could be observed in both hemispheres, mostly in the equatorial region.
From the sinusoidal off-limb curve in Figure~\ref{F-offlimbfans}, one can estimate the corresponding rotation rate of the feature. The rotation rates for different stripes (eight in total) show a large range from 10\,degree\,day$^{-1}$ to 15\,degree\,day$^{-1}$, with an average value of 12.45\,degree\,day$^{-1}$. The large variation in the speeds is also due to the EUV intensity variations resulting from the integration along the line-of-sight of different dynamical coronal features.

\subsection{The Evolution of the Average Solar Corona Based on EUV Intensity Changes \label{subsec:avgcorona}}

To better quantify the EUV coronal evolution in brightness and extent, we measured the mean brightness of
each image [data numbers: DN] in different sectors, at different distances from the Sun center and we compared them to the corresponding on-disk brightness, LYRA irradiance and international sunspot number (ISN: SIDC - \url{sidc.oma.be/silso/datafiles}). An example of one sector is shown in the Appendix~\ref{S-sectors}.

\subsubsection{The Evolution of the On-Disk Average Solar Corona}

\begin{figure}
\centering
\includegraphics[width=\textwidth]{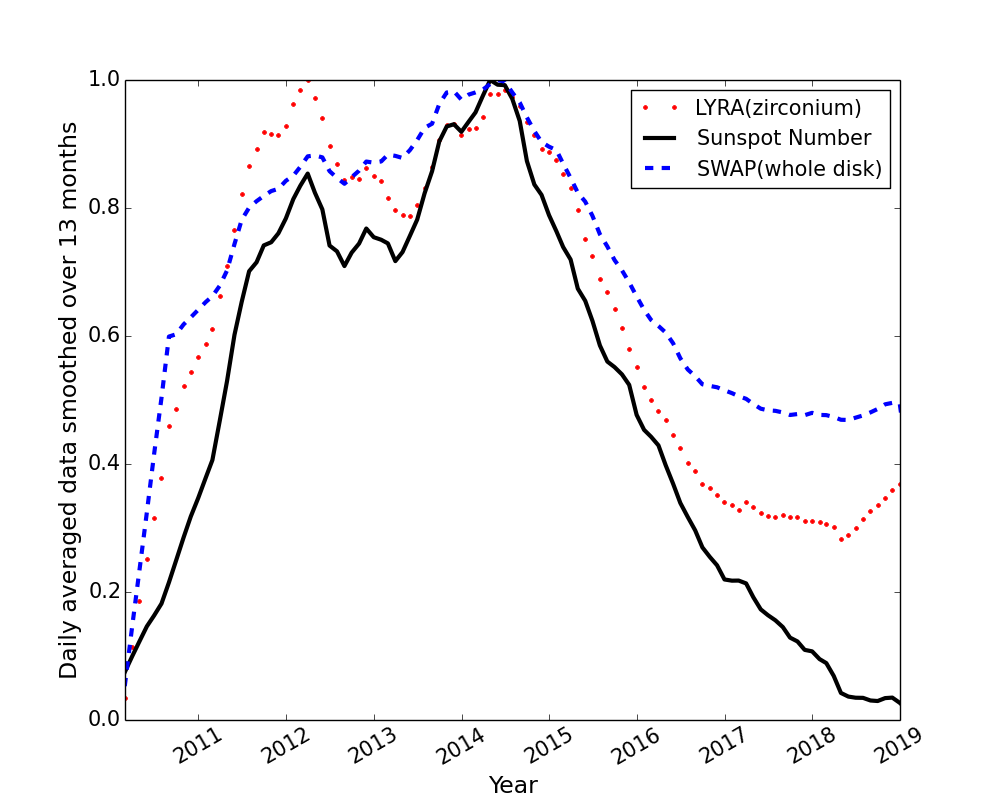}
\caption{Sunspot number (\textit{black-continuous}), SWAP EUV disk brightness (\textit{blue-dashed}) and LYRA zirconium irradiance (\textit{red-dotted}) as a function of time for the period February 2010\,--\,June 2019.}
\label{F-diskevol}
\end{figure}

Figure~\ref{F-diskevol} shows the evolution of the whole disk EUV corona brightness observed by SWAP (in  blue-dashed) for the period February 2010\,--\,June 2019. 
Overplotted in black-continuous is the sunspot number  and in red-dotted is the LYRA zirconium irradiance. In this figure, all of the data (SWAP, LYRA, and Sunspot Number) are averaged over each  month, smoothed with the \textsf{convolve} Python function from the \textsf{Numpy} Package, over 13-month boxes, and normalized using Equation~\ref{E-normalization}.

\begin{equation} \label{E-normalization}
z_{i}=\frac{x_{i} - \min(x)}{\max(x) - \min(x)}
\end{equation}

In Equation~\ref{E-normalization}, $x_{i}$ represents the data point at a given time, $x$ is the data vector ($x=(x_{1},...,x_{n})$) and $z_{i}$ is the $i$th normalized data.

The sunspot number exhibits two maximum peaks (the second higher than the first), the first one taking place in February 2012 and the second one in March 2014.
We observe that the SWAP on-disk brightness (blue line) follows the sunspot number trend (black line). The correlation between LYRA and sunspot number data is 0.97, between SWAP on-disk brightness and sunspot number is 0.94 and between LYRA and SWAP on-disk brightness is 0.97. The three time series are very well correlated. We also compared the evolution of the entire EUV coronal brightness observed by SWAP (on-disk + off-limb) with LYRA signal and sunspot number. In this case the Pearson correlation coefficients are: LYRA--ISN: 0.97, SWAP--ISN: 0.95, and LYRA--SWAP: 0.97. The Pearson correlation coefficient is a measure of the linear correlation between two variables.

\begin{figure} 
	\centering
	\includegraphics[width=\textwidth]{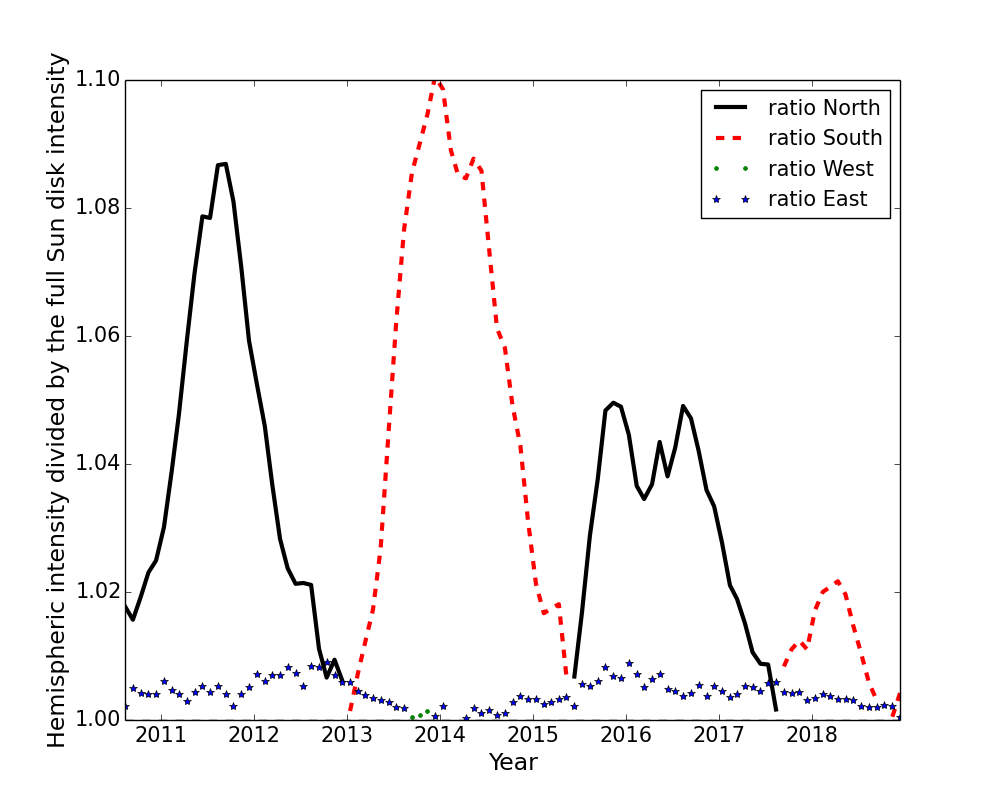}
	\caption{Ratio between the hemispheric SWAP averaged signal and the SWAP whole disk averaged signal smoothed over 13-month period: The northern hemisphere ratio (\textit{black-continuous}), the southern hemisphere ratio (\textit{red-dashed}), the eastern hemisphere ratio (\textit{blue-stars}) and the western hemisphere ratio (\textit{green-dotted}).}
	\label{F-swapratio13}
\end{figure}

Figure~\ref{F-swapratio13} shows the evolution of the ratio between the on-disk averaged EUV brightness observed for the distinct SWAP hemispheres and the whole on-disk averaged brightness in [DN\,s$^{-1}$\,pix$^{-1}$] over the period from February 2010 to June 2019. We used the daily averaged data, and we smoothed each of them with a 13-month boxcar filter. Note that the ratios can be higher than unity because of the relationship between the averaged values: Average(North) + Average(South) = 2 Average(Total Sun).

\begin{figure}
\centering
\includegraphics[width=0.95\textwidth]{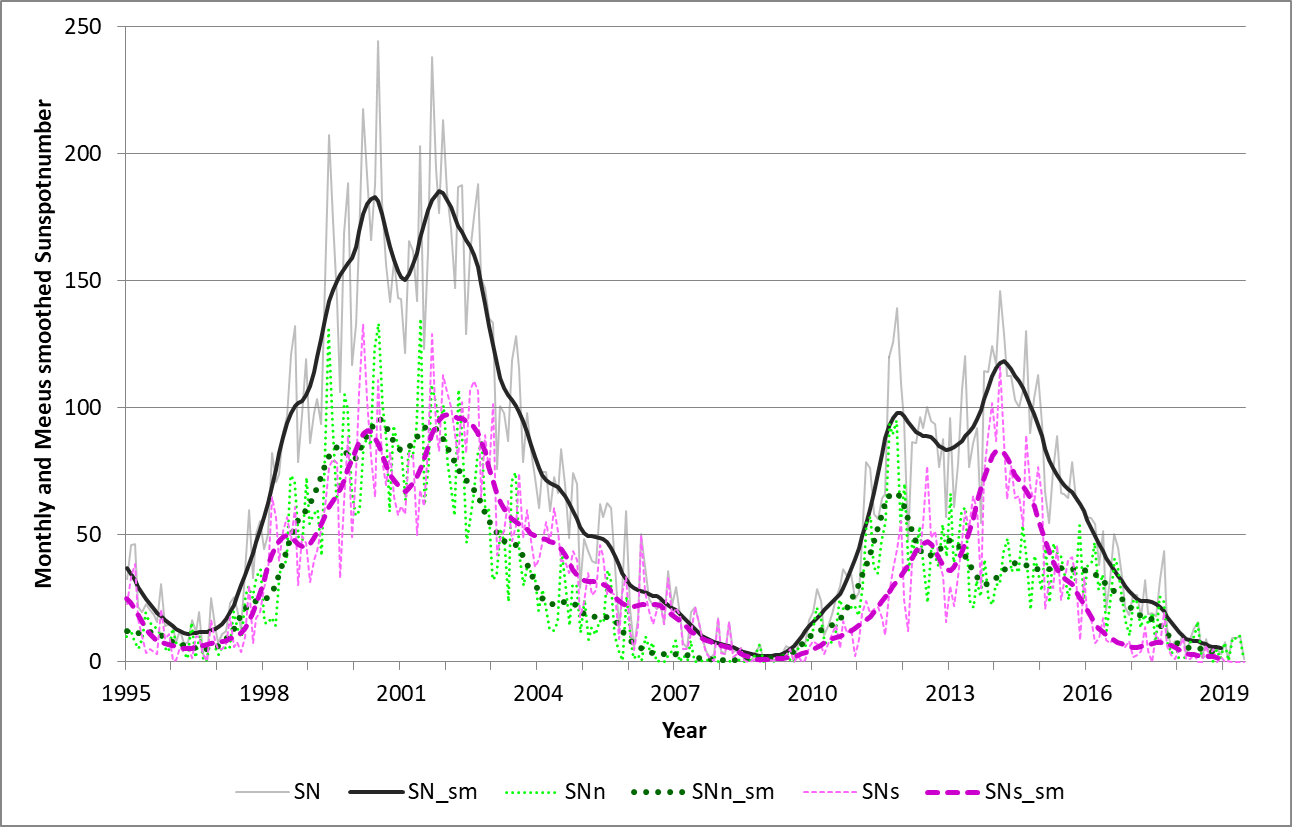}
\caption{Evolution of the total and hemispheric sunspot number for SC23 and SC24 (period 1995\,--\,2019). The raw monthly values are respectively in \textit{dotted light-green} (North), \textit{dashed pink} (South) and \textit{gray} (total), the Meeus smoothed values are respectively in \textit{dotted dark-green} (North), \textit{dashed purple} (South), and \textit{dark-gray} (total). It is clear that the first SC24 maximum was due to increased activity on the northern solar  hemisphere (smoothed ISNn = 66.4 in November 2011), whereas the second, true, maximum was the result of enhanced activity on the southern hemisphere (smoothed ISNs = 83.3 in February 2014).} \label{F-hemisphsn}
\end{figure}

We can see that the ratio for the southern hemisphere SWAP signal is below the northern hemispheric ratio the majority of the time except for the period 2013\,--\,mid 2015 and over a few other small periods of time. 
This is explained by the fact that, as observed in Figure~\ref{F-hemisphsn}, the first peak of the SC24 (in November 2011) is dominated by the northern hemisphere, while the second peak (in March 2014) is dominated by the southern hemisphere\footnote[4]{Note that the values derived from Figure~\ref{F-hemisphsn} are distinct from the ones derived from Figure~\ref{F-diskevol} due to different smoothing procedures.}. However, the ratio for the northern hemisphere (black-continuous curve in Figure~\ref{F-swapratio13}) dominates over the southern hemisphere (red-dashed curve in Figure~\ref{F-swapratio13}) throughout our studied period, except for the second maximum peak when the southern hemisphere dominates. 

On the other hand, the ratio for the eastern hemisphere (blue-stars) and the western hemisphere (green-dotted) are very similar (close to unity). The eastern ratio is a little higher than the western hemisphere ratio. 

It is clear that the eastern hemisphere brightness dominates over the western one all over the studied period, except a small interval at the end of 2013, and the northern hemisphere dominates over the southern one except for the periods January 2013\,--\,May 2015 and the end of 2017\,--\,June 2019. 

\subsubsection{The Evolution of the Off-Limb Average Solar Corona Emission}

\begin{figure}
	\centering
	\includegraphics[width=\textwidth]{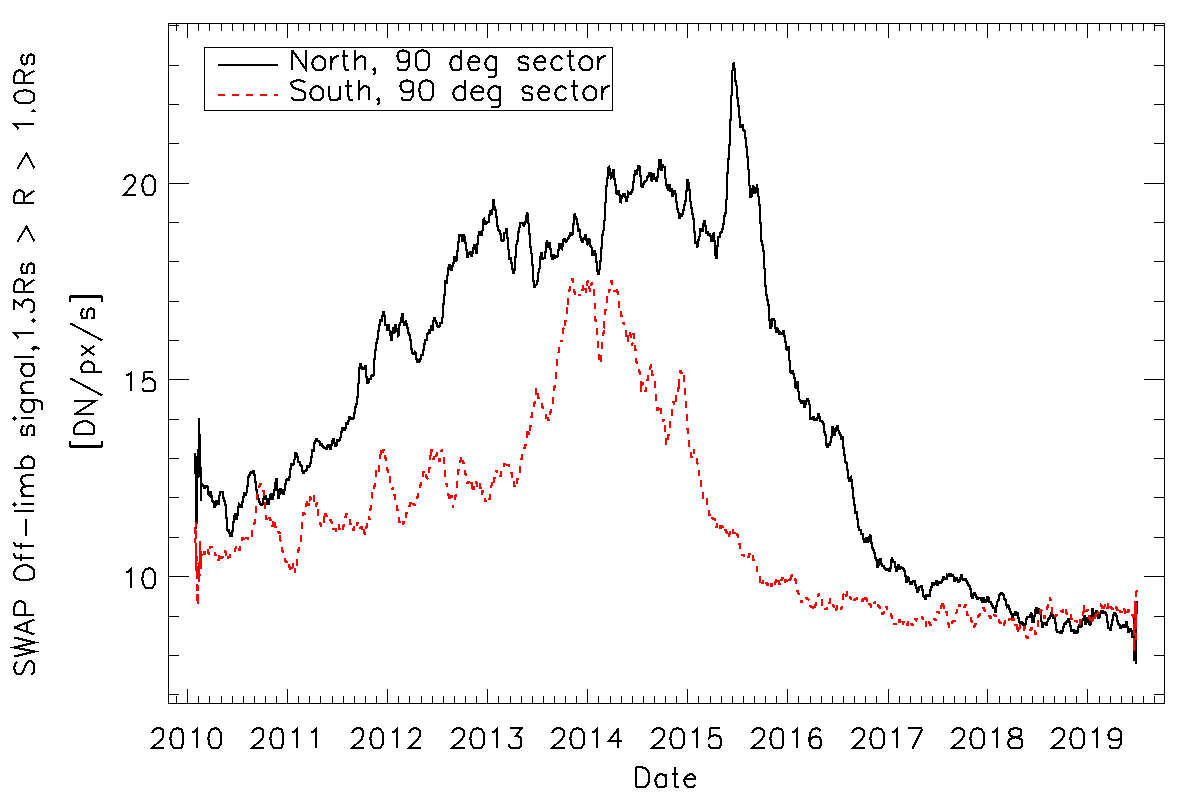}
	\caption{The evolution of the average solar corona (smoothed over 30\,days) at the South Pole (\textit{red-dashed}) and the North Pole (\textit{black}) (sector of 90\,degrees centered on the pole), from 1\,R$_{\odot}$ to 1.3\,R$_{\odot}$.} \label{F-avgpoles90}
\end{figure}

Figure~\ref{F-avgpoles90} shows the average solar coronal emission at the Poles as a function of time. The average is taken in sectors of 90 degrees centered on the North Pole (black) and South Pole (red), from 1\,R$_{\odot}$ to 1.3\,R$_{\odot}$ (see also Figure~\ref{F-synopticonoff} for the selection of the sectors).

\begin{figure}
	\centering
	\includegraphics[width=\textwidth]{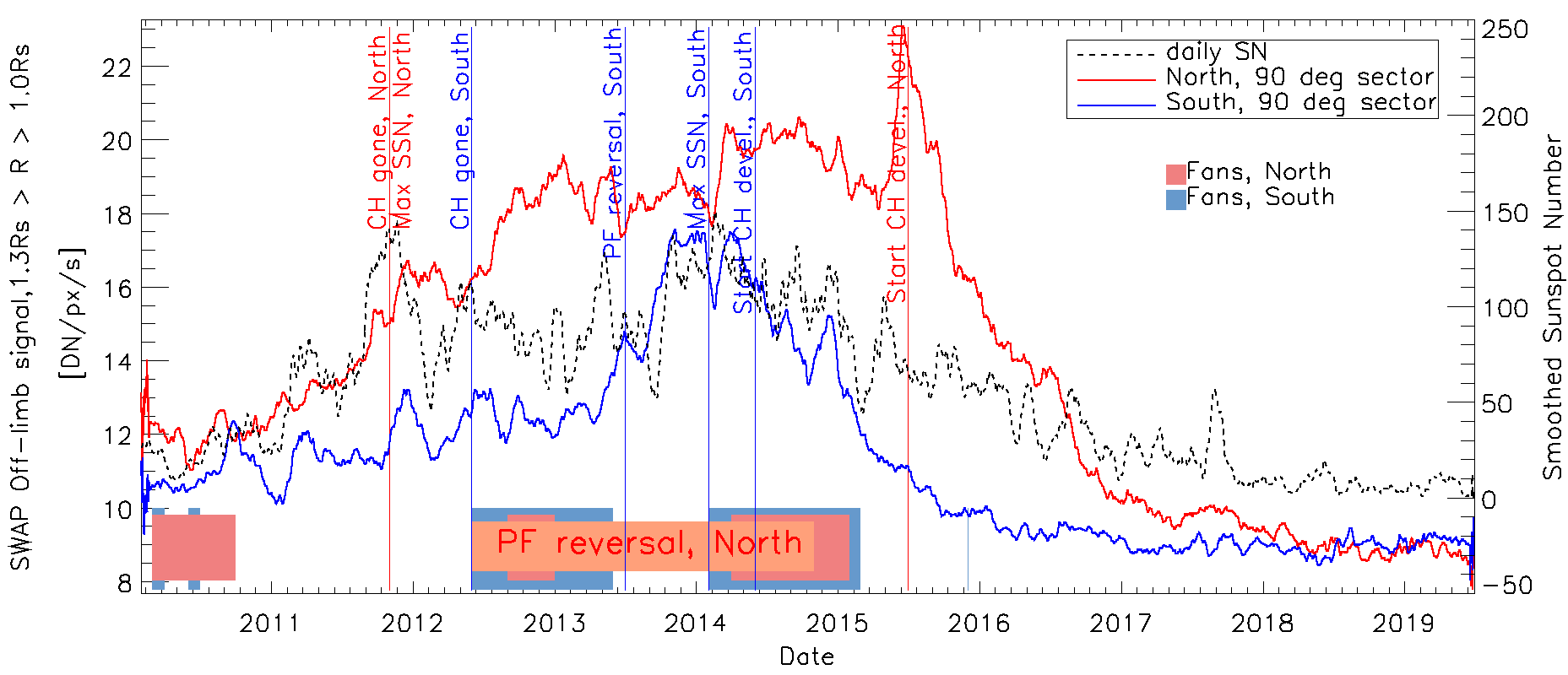}
	\caption{Overview plot. The evolution of the 30-day averaged solar corona at South Pole (\textit{blue}) and North Pole (\textit{red}) in a sector of 90\,degrees centered on the Pole, from 1\,R$_{\odot}$ to 1.3\,R$_{\odot}$. The evolution of the 30-day averaged sunspot number is shown \textit{in black}. The \textit{light blue background} indicates a period when fans and (pseudo)streamers are seen near the South Pole, and the \textit{pink background} indicates a period when fans and (pseudo)streamers are seen near the North Pole. Different other features at North (\textit{red}) and South (\textit{blue}) are indicated by the \textit{vertical lines}. The time period of northern polar field reversal is indicated by the \textit{orange background}.}
	\label{F-overviewplot}
\end{figure}

One can see a secondary peak (spike) on the descending phase of the SC24. For the northern sector the peak is more pronounced compared with the southern one and it takes place around June 2015. The smaller peak (spike) at southern sector is observed around December 2014. The ISN dataset, smoothed according to the Meeus formula, has two maxima respectively in November 2011 (98.0) and in March 2014 (118.2)--see Figure~\ref{F-hemisphsn}. Note that these values are distinct from the ones derived from Figure~\ref{F-diskevol} due to different smoothing procedures. The hemispheres peaked respectively in November 2011 (North, 66.4) and in February 2014 (South, 83.3); see also Table~\ref{T-polphenom} and Figure~\ref{F-overviewplot}. 

\begin{figure}
\centering
\includegraphics[width=0.49\textwidth]{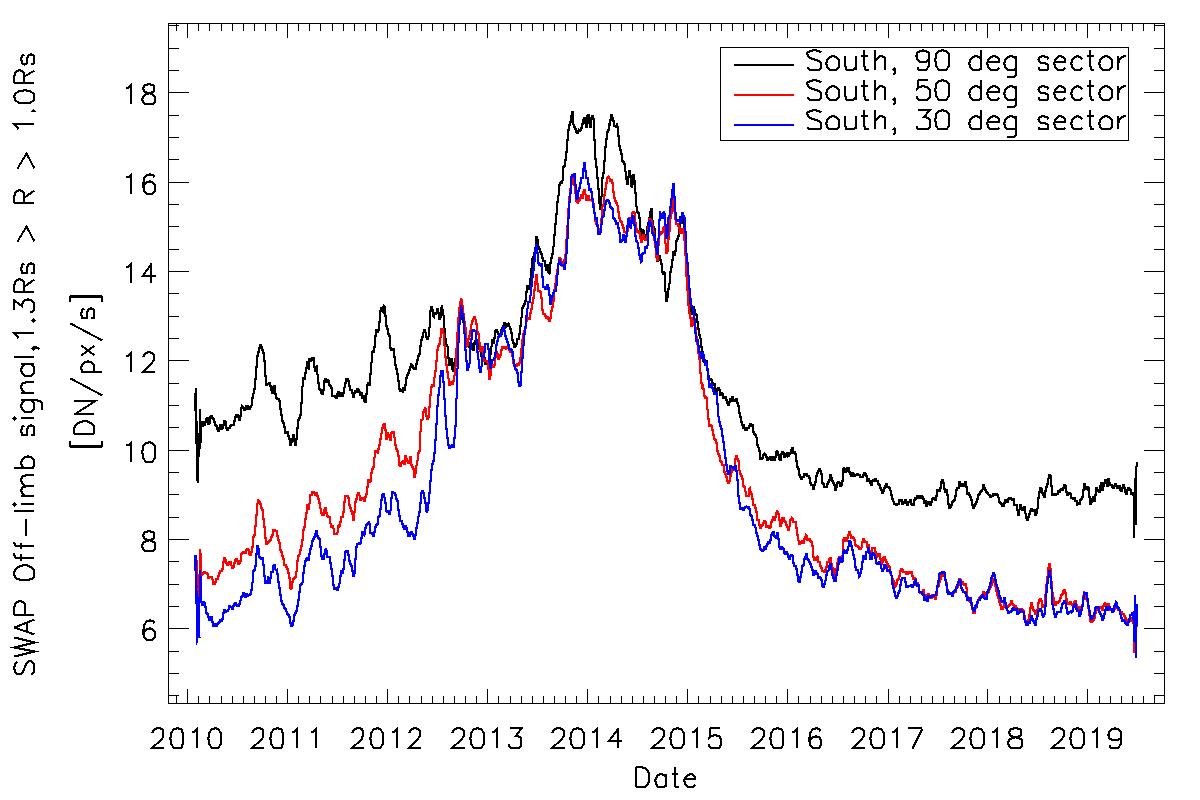}
\includegraphics[width=0.49\textwidth]{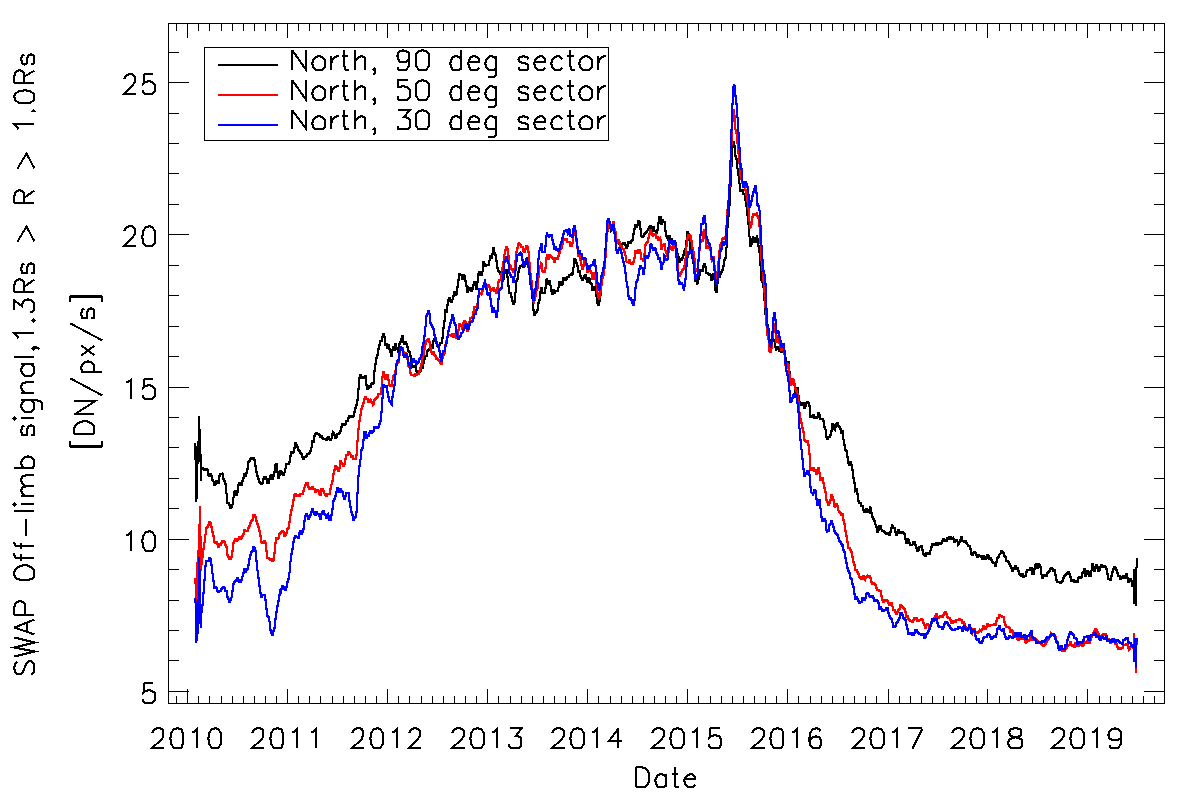}
\caption{\textit{Left panel}: The evolution of the 30-day averaged solar corona at the South Pole (sector of 90\,degrees centered on the Pole \textit{in black}, sector of 50\,degrees centered on the Pole \textit{in red} and sector of 30\,degrees centered on Pole \textit{in blue}), from 1\,R$_{\odot}$ to 1.3\,R$_{\odot}$. \textit{Right panel}: The same as the left panel, but for the North Pole.} \label{F-avgpoles}
\end{figure}

Similar to Figure~\ref{F-avgpoles90}, we plot in Figure~\ref{F-avgpoles} the average solar corona at the Poles in sectors of 90 (black), 50 (red) and 30 (blue) degrees centered on the North Pole (right panel) and the South Pole (left panel), from 1\,R$_{\odot}$ to 1.3\,R$_{\odot}$. From Figure~\ref{F-avgpoles} it is clear that the two peaks present in all sectors are not dependent on the selection of the sector width.

\begin{figure}[ht!]
\centering
\includegraphics[width=0.95\textwidth]{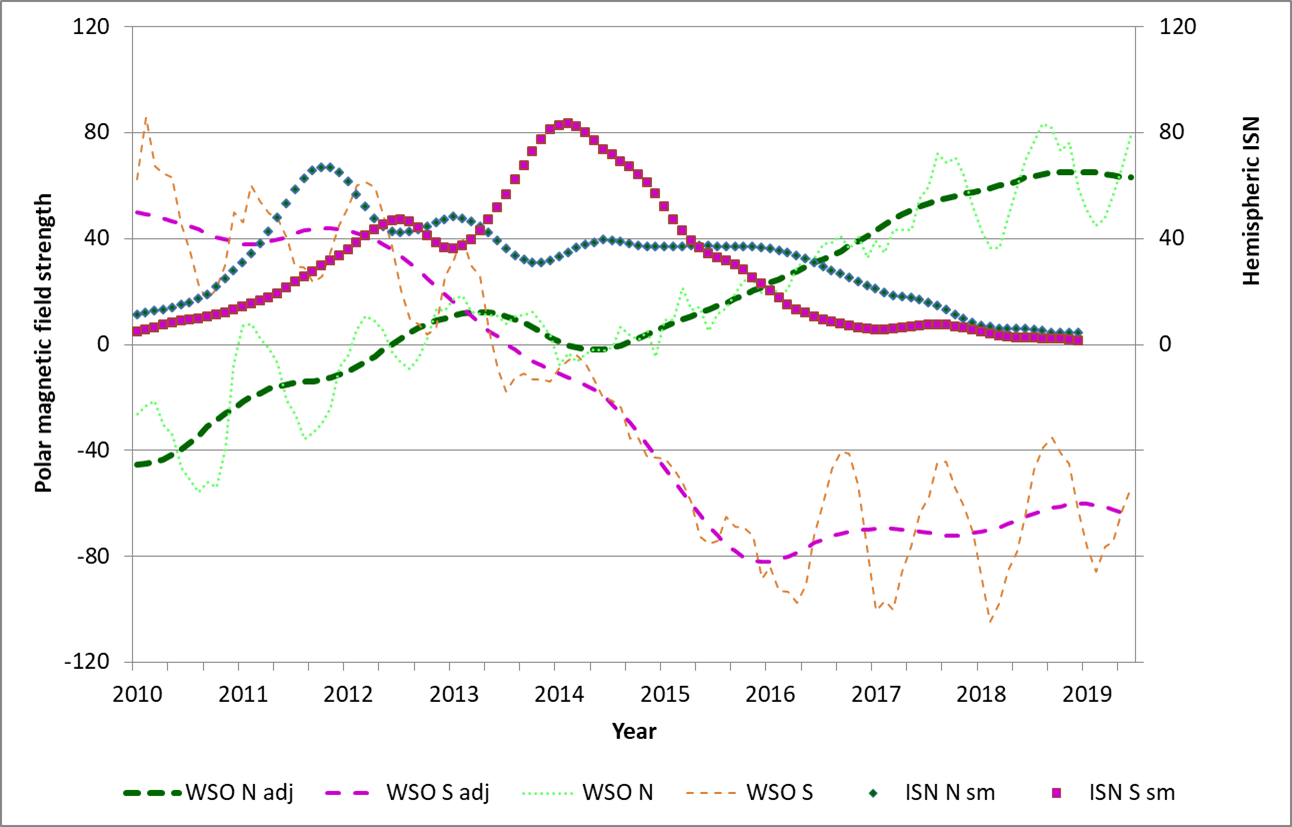}
\caption{Polar magnetic field strength (WSO) and sunspot numbers (SILSO) per solar hemisphere. The magnetic-field strength is expressed in 0.01 Gauss, with raw (\textit{thin-dashed line}) and adjusted values (\textit{thick-dashed line}) for the northern and southern solar pole in respectively \textit{light / dark green} and \textit{orange / purple}. The Meeus smoothed ISN (dimensionless) is expressed by the \textit{blue-greenish diamonds} and the \textit{purple squares} representing respectively the northern and southern hemisphere. } \label{F-mfpoles}
\end{figure}

To further investigate the presence of these spikes at the Poles on the descending phase of the SC24, we plot the magnetic field at the Poles in Figure~\ref{F-mfpoles}. We see that the reversal at the South Pole was in July 2013, while for the North Pole it took longer, from June 2012 to November 2014 (see also \citealt{Janardhan2018}). The spike at North in SWAP is observed three years after the beginning of the northern polar magnetic field reversal, and 7 months after the end of the northern polar magnetic field reversal. In South, the spike is observed 17 months after the southern polar magnetic field reversal. A summary of all polar phenomena observed during the SC24 is outlined in Table~\ref{T-polphenom} and Figure~\ref{F-overviewplot}.
Magnetic-field timings come from Figure~\ref{F-mfpoles} and the hemispheric maxima were read from the Meeus smoothed sunspot number.

\begin{table}
\caption{Summary of polar phenomena in SC24. Time is expressed in year.month. CH = Coronal Hole and PF = Polar Field.} \label{T-polphenom}
\begin{tabular}{c|cc}
	\hline
Time & North & South \\
   \hline
2010.03--2010.04 &  & Coronal fan \citep{Talpeanu2016} \\
2010.03--2010.10 & 3 Coronal fans \citep{Talpeanu2016} &  \\  
2010.06--2010.07 &  & Coronal fan \citep{Talpeanu2016} \\ 
2011.11 & Polar CH gone &  \\
2011.11 & Max. sunspot number &  \\
2012.06 &  & Polar CH gone \\
2012.06 & Starts PF reversal &  \\
2012.06--2013.06 &  & Pseudostreamer \citep{Seaton2013b} \\
2012.07--2012.10 &  & 2 Coronal fans \citep{Talpeanu2016} \\
2012.09--2013.01 & Coronal fan \citep{Talpeanu2016} &  \\
2013.01 &  & Coronal fan \citep{Talpeanu2016}  \\
2013.05 &  & Pseudostreamer \citep{Rachmeler2014} \\
2013.07 &  & PF reversal \\
2014.02 &  & Max sunspot number \\
2014.02--2015.03 & & Pseudostreamer \citep{Guennou2016}\\
2014.04--2015.02 & 3 Coronal fans \citep{Talpeanu2016} &  \\
2014.06 &  & Start polar CH development \\
2014.11 & End PF reversal &  \\
2014.11 &  & Pseudostreamer \\
2014.12 &  & Peak (spike) SWAP \\
2015.06 & Peak (spike) SWAP &  \\
2015.07 & Start polar CH development &  \\
2015.12 &  & Pseudostreamer \\
2017.06--2017.10 & Max. magnetic field  &  \\
\end{tabular}
\end{table}

These SWAP ''peaks or spikes'' in Table~\ref{T-polphenom} seem to be associated with the start of the development of the (polar) CHs. This spike in intensity seems to come from the fuzzy emission near the Poles, as the peak remains prominently visible in Figure~\ref{F-avgpoles}, even when the sector arc became smaller (from 90$^{\circ}$ to 30$^{\circ}$). 
From the synoptic maps, the fuzzy emission near the Poles (meaning no CHs) is observed to last from December 2011 to April 2015 at the North Pole, and from December 2013 to August 2014 at the South Pole.

\section{Discussion} \label{S-discussions}
We studied the long-term evolution of the solar corona over the SC24, by analyzing the EUV intensity variations in terms of the average solar-rotation and by looking at the magnetic-field evolution. We studied the association between the coronal EUV brightness, the sunspot number, and the solar irradiance. We were also able to monitor and discuss large-scale structures like fans and/or (pseudo)streamers.

\subsection{Solar-Cycle Evolution}

The positions of sunspots throughout a solar cycle are often revealed in the classic butterfly diagram, with sunspots first appearing at high latitudes and then emerging progressively closer to the Equator throughout the cycle evolution. Such a pattern can also be observed in Figure~\ref{F-swapsynoptic}, which shows the SWAP synoptic maps (for central meridian) at different times during SC24. The bright features on the maps indicate ARs, fans, streamers, pseudostreamers and/or bright points and the dark features indicate CHs and/or filaments.

The polar CHs are well visible towards the minimum of solar activity (see lower panels of Figure~\ref{F-swapsynoptic} showing the on-disk corona for CR 2195 (September--October 2017) and CR 2218 (June 2019)). They are completely missing at maximum of solar activity (see CRs 2147 and 2157 maps, corresponding to year 2014, in Figure~\ref{F-swapsynoptic}).

\cite{McIntosh2003} first noted that both the polar crown of filaments and sunspots begin around the same period: early in a cycle around 40$^{\circ}$ latitude. Then they diverge with the polar coronal filaments moving poleward and sunspots and active regions moving equatorward. 
Polar filaments drift to the Poles before the polar-field reversal near solar-cycle maximum (so-called rush to the poles;  \citealt[see, \textit{e.g.},][]{Lockyer1931, Hyder1965, Hansen1975}) and thus represent a characteristic feature of the highly active Sun.

\subsection{Dynamics of the Solar Features in Terms of Solar Rotation}
Differential rotation of the Sun was studied intensively using different types of data: sunspots \citep[see, \textit{e.g.},][]{Zhang2013, Li2014, Ruzdjak2017}, solar filaments \citep[see, \textit{e.g.},][]{Glackin1974, Adams1977, Dzaparidze1992, Brajsa1997, Gigolashvili2013}, bright points \citep{Karachik2006, Hara2009}, coronal holes \citep{Adams1976, Shelke1985, Insley1995, Hiremath2013, Bagashvili2017, Krista2018}, magnetic field \citep{Shi2013, Shi2014, Suzuki2014, Lamb2017, Badalyan2018, Imada2018, Xie2018}. 
Off-limb solar-corona rotation was studied using coronagraphic white-light images as well as line emission corona data \citep[see, \textit{e.g.},][]{Stenborg1999, Giordano2008, Mancuso2011, Morgan2011, Mancuso2012, Mancuso2013, Bhatt2017}.

It was found that the Sun rotates more differentially at the minimum than at the maximum of activity during the epoch 1977\,--\,2016 \citep{Ruzdjak2017}.

The solar-surface-rotation rate at the Equator shows a decrease since Cycle 12 onwards, given by about 1\,--\,1.5 $\times$10$^{-3}$\,degree\,day$^{-1}$\,year$^{-1}$, while the rotation rate averaged over latitudes 0$^{\circ}$\,--\,40$^{\circ}$  does not show a secular trend of statistical significance \citep{Li2014}. 

\cite{Shi2013}, through a cross-correlation analysis of the Carrington synoptic maps of solar photospheric magnetic fields from February 1975 to October 2012, found that the sidereal rotation rates decrease from the Equator to mid-latitude and reach their minimum values of about 13.16\,degree\,day$^{-1}$ at $\pm$53$^{\circ}$ latitude, then increase toward higher latitudes. They also showed that the rotation rates seem to decrease at the beginning of a solar cycle, and within the descending phase of a solar cycle, they increase and then decrease again. This solar-cycle variation is visible more clearly at higher latitudes.
When  magnetic fields are weaker, one can expect a more pronounced differential rotation yielding a higher rotation velocity at low latitudes on average \citep{Li2014}.

From our study on the dynamics of the solar features as the Sun rotates, no conclusive remarks regarding the SC evolution can be derived. On average, we estimated rotation rates of around 15\,degree\,day$^{-1}$ at latitudes of -15$^{\circ}$, 0$^{\circ}$, and 15$^{\circ}$ throughout the SC24. Note that the bright stripes used to estimate the average solar rotation may indicate a superposition of ARs or bright points or large scale structures at different latitudes and altitudes, and as a consequence, the errors introduced are big (values up to 2.4\,degree\,day$^{-1}$). This may also be the reason that we do not see any difference between rotation rates at different latitudes--see Table~\ref{T-rotation}.
This is in agreement with other studies found in literature: \cite{Hara2009} showed that the evaluated rotation rate of X-bright points (XBPs) for long-lived XBPs (lifetime larger than eight hours) follows the rotation rate of the photospheric magnetic fields, and short-lived XBPs rotate slower than long-lived XBPs.
\cite{Karachik2006} demonstrated that coronal features situated at the same heliographic coordinates but different heights in the corona may exhibit different rotation rates. 
Supergranules rotate faster than sunspots, and younger sunspots rotate faster than the older ones \citep[see, \textit{e.g.}, the review by][]{Beck2000}.  

We found that average rotation rates at latitudes S15 and N15 are close to the expected values \citep{Snodgrass1990}.

\subsection{Polar Magnetic Field Reversal}

Using McIntosh Archive data, \cite{Webb2018} noted that the polar field reversal process is manifested by two features: i) the end of the filament rush to the poles and ii) the disappearance of the polar CH before this and, some time later, the reversal of the polarity of the CH closest to the Pole.

Usually when the polar field is minimum (values close to zero), the field in the sunspot belt zone is maximum \citep{Janardhan2018}. From Figure~\ref{F-mfpoles}, it can be seen that the polar photospheric magnetic fields have already approached the minimum at the southern pole by mid-2013, while at the northern pole it has a prolonged minimum period from mid-2012 to the end of 2014.

From the SWAP data we have observed a secondary peak (at both North and South Pole) on the descending phase of the SC24 (see Figure~\ref{F-overviewplot}). These peaks seem to be related to the start of the development of the (polar) coronal holes. They seem to come from the fuzzy emission near the Poles, as the peak remains prominently visible in Figure~\ref{F-avgpoles}, even when the sector (where the average intensity was calculated) became smaller (from 90$^{\circ}$ width to 30$^{\circ}$ width).

In Figure~\ref{F-mfpoles} we show the polar magnetic-field strength per solar hemisphere. It is observed that the field reversal at the South Pole takes place in July 2013, seven months before the maximum sunspot number for that hemisphere (February 2014). In the northern hemisphere, the sunspot number reaches its maximum in November 2011, whereas the field
reversal takes place between June 2012 and November 2014 (see also \citealt{Janardhan2018}). 

A well-defined polar coronal hole was present in the southern hemisphere by October 2015, approximately two years after the field reversal in the South. The large polar coronal hole in the northern hemisphere has not fully developed in October 2015 while it had fully matured in October 2016, a year later  (\citealt{Golubeva2016}). The dates for the southern coronal hole are different from the dates displayed in Table~\ref{T-chs} in the case of Observers 2 and 3, confirming once more the subjectivity in observing the appearance and disappearance of CHs from different data sources and by different observers.

\cite{Petrie2017} used NSO/Kitt Peak synoptic magnetograms covering Cycles 21\,--\,24, and they concluded that in the most abrupt cases of polar field reversal, a few activity complexes (systems of active regions) were identified as the main cause. This is due to the poleward transport of large quantities of decayed trailing-polarity flux from these complexes, which was found to correlate well in time with the abrupt polar-field changes. The Cycle 21 and 22 polar reversals were dominated by only a few long-lived complexes whereas the Cycle 23 and 24 reversals were the cumulative effects of more numerous, shorter-lived region complexes.
In our case, for the southern reversal in SC24, there were three such regions that contributed significantly with new magnetic flux, and it was a lack of such ARs in the northern hemisphere that delayed a firm polar-field reversal at the northern solar pole.
 
 \subsection{Large-Scale Coronal Features}
 Coronal pseudostreamers, which separate like-polarity coronal holes, do not have current sheet extensions, unlike the familiar helmet streamers that separate opposite-polarity holes. Both types of streamers taper into narrow plasma sheets that are maintained by continual interchange reconnection with the adjacent open magnetic field lines. 
 
 \cite{Wang2007} showed that the principal morphological difference between the pseudostreamers and the streamers is that the helmet streamer cusps can be seen above the LASCO-C2 occulter (2\,R$_{\odot}$), whereas only the long stalks of the pseudostreamers are visible above 2\,R$_{\odot}$.
 
 \cite{Guennou2016} reported on a coronal pseudostreamer in the southern polar region of the solar corona that was visible for approximately a year starting in February 2014. Following that, the pseudostreamer gradually shrinks until its disappearance in March 2015, at which point the southern polar coronal hole once again dominates the Pole.
 The pseudostreamer studied by \cite{Guennou2016} was captured in the off-limb synoptic maps (see Figure~\ref{F-offlimbimg}) as bright inclined stripes. The stripe from lower-left to middle-right of the map indicates a front-disk transit of the pseudostreamer and the stripe from middle-right to upper-left indicates a behind-disk transit of the feature. The persisting features at the South Pole can be seen in the left panel of Figure~\ref{F-offlimbimg} as a bright vertical band. These are probably bright spicules which are always present at those distances (1.1\,R$_{\odot}$) at the poles. Some are still visible at 1.3\,R$_{\odot}$ (middle panel of Figure~\ref{F-offlimbimg}) but they disappear completely at 1.6\,R$_{\odot}$ (right panel of Figure~\ref{F-offlimbimg}).
 
Other large-scale coronal features, besides streamers and pseudostreamers, are the EUV fans. These are large fan-like structures that extend to extremely large distances above the solar surface, up to several solar radii. They are seemingly open features, connecting to the solar surface at footpoints extending away from the Sun in the other direction. Near their footpoints they appear almost radial, but above that, they bend around large closed loops.
\cite{Talpeanu2016} studied the fans observed by SWAP in the period between March 2010 and July 2010, and between July 2012 and October 2014. She found that all footpoints remain in the latitude interval between [-40$^{\circ}$, + 40$^{\circ}$], showing correlation to the active region bands, even though in only half of the cases, the fans were associated with large active regions. For almost all the fans the identified footpoints remain in the same magnetic domain and do not cross a polarity inversion line, meaning that they are unipolar. Almost half of the footpoints were near coronal holes but none of the	footpoints was located inside a coronal hole.
A long-lived fan (more than 11 CRs) studied by \cite{Talpeanu2016} can also be seen to zigzag across Figure~\ref{F-offlimbfans}. When the fan is rotating in front of the solar disk, the stripes span the map from lower-left to upper-right, and when the fan is rotating behind-disk the stripes span the map from lower-right to upper-left. For a few CRs the fan is visible even at 1.6\,R$_{\odot}$ (not shown in the figure). 

From the sinusoidal off-limb curve of the fan in Figure~\ref{F-offlimbfans}, one can determine the corresponding rotation rate of the feature, and from that the latitude at which the footpoint of the magnetic structure is anchored. The rotation rates estimated for different stripes visible in Figure~\ref{F-offlimbfans} range from 10 degree\,day$^{-1}$ to 15\,degree\,day$^{-1}$, with an average value of 12.45\,degree\,day$^{-1}$. This large range of values may indicate that the fan is not rigidly anchored to its footpoints and/or other phenomena in the solar corona contribute to its rotation rate. The main cause may have its origins in the fact that the EUV intensity variations in SWAP images result from the integration of the signal of all features along the line-of-sight.

\section{Conclusions} \label{S-conclusions}

Using level-1 SWAP images, level-3 LYRA irradiance time series, and the version 2.0 sunspot data, we studied the evolution of the solar corona throughout SC24 (from 2010 to 2019). 

Our results can be summarized as follows:
        \begin{itemize}
            \item From SWAP synoptic maps (see Figure~\ref{F-swapsynoptic}) one can see when the northern or southern hemisphere is more active and when polar CHs are present or disappear.
                \begin{itemize}
                   \item More ARs started to appear (starting in the northern hemisphere) in February 2011 and they became less frequent beginning in December 2016, reaching a very low number from September 2017 onward, indicating the passage from solar minimum to maximum and back again.
                   \item In the northern hemisphere the polar CHs were present from February 2010 to October 2011, with some short intermittent periods. No CHs were observed between November 2011 and June 2015. They started to develop again in July 2015 and they remained visible until June 2019 (end of our dataset).
                   \item At the South Pole the CHs were present from February 2010 to May 2012, with some intermittent periods. No CHs were observed between June 2012 and May 2014. They started to develop again in June 2014 and they remained visible until June 2019 (end of our dataset).
                   \item Note that there is quite some discrepancy, depending on the observer and data used, in identification of polar coronal holes. This is why the numbers above are approximations. Comparison with automated CHs detection algorithms as CHIMERA also suggest that there are open field lines not well observed in the SWAP synoptic maps and/or EUV images.
                \end{itemize}
           \item From the EW synoptic maps (see Figure~\ref{F-ewswapsynoptic2157}) one can calculate the rotation of the solar coronal features at different latitudes and estimate when these features are formed or disappear. 
                \begin{itemize}
                   \item The average rotation speed, for bright features observed between latitudes of -40$^{\circ}$ and +40$^{\circ}$ is approximately 14 deg/day, with an estimated maximum error of 2.4\,degree\,day$^{-1}$.
                   \item The average rotation rate of the features at latitudes of +15$^{\circ}$, 0$^{\circ}$ and -15$^{\circ}$ is around 15\,degree\,day$^{-1}$ throughout the whole SC24. The average rotation rates of the bright features located at S15 and N15 (14.5\,degree\,day$^{-1}$) are close to the expected values of the photospheric rotation rates \citep{Snodgrass1990}.
              \end{itemize}  
       \end{itemize}

     \begin{itemize}
     	\item From the off-limb synoptic maps (see Figure~\ref{F-offlimbfans}) one can follow the evolution of large-scale coronal features (fans, streamers, pseudostreamers).
     	   \begin{itemize}
               \item Large-scale off-limb structures were visible from around March 2011 to around March 2016, meaning that they were absent at the minimum phase of the solar cycle.
               \item  A fan in the northern hemisphere was seen to persist for more than 11 CRs (February 2014 to March 2015) and was observed out to 1.6\,R$_{\odot}$. Its rotation rates range from 10\,degree\,day$^{-1}$ to 15\,degree\,day$^{-1}$, with an average value of around 12\,degree\,day$^{-1}$.
           \end{itemize}  
      \end{itemize} 
  
   \begin{itemize}  
 	  \item SWAP on-disk average brightness follows the sunspot number trend.
      \item The linear correlation between LYRA and sunspot number data is 0.97, between SWAP on-disk brightness and sunspot number is 0.94 and between LYRA and SWAP on-disk brightness is 0.97.
      \item The ratio between the SWAP southern hemisphere averaged brightness and the whole on-disk averaged brightness is below the northern ratio in majority of the time except for the period 2013--mid-2015 and few other small periods. This is in accordance with the fact that the first peak of the SC24 (observed on November 2011) is dominated by the northern hemisphere, while the second peak (observed on March 2014) is dominated by the southern hemisphere.	
      \item The ratio between the SWAP eastern hemisphere averaged brightness and the whole on-disk averaged brightness is very similar to the corresponding ratio for the western hemisphere (close to unity). Eastern hemisphere emission dominates over the western one all over the studied period, except a small time interval at the end of 2013.     
   \end{itemize}  
   
   \begin{itemize}
      \item A sharp peak was observed in the North Pole SWAP intensity in June 2015 (after the SC24 maximum phase); see Figure~\ref{F-overviewplot}. 
      \item A smaller peak was observed in the South towards the end of 2014. 
      \item The peaks described above seem to be associated with the start of the development of the (polar) coronal holes. 
   \end{itemize}

\begin{acks}

We acknowledge the use of SWAP, LYRA, SILSO, and WSO data. SWAP is a project of the Centre Spatial de Li\`ege and the Royal Observatory of Belgium funded by the Belgian Federal Science Policy Office (BELSPO). LYRA is a project of the Centre Spatial de Li\`ege, the Physikalisch-Meteorologisches Observatorium Davos and the Royal Observatory of Belgium funded by the Belgian Federal Science Policy Office (BELSPO) and by the Swiss Bundesamt f{\"u}r Bildung und Wissenschaft. Sunspot data source: WDC-SILSO, Royal Observatory of Belgium, Brussels. Wilcox Solar Observatory data used in this study were obtained via the website \url{wso.stanford.edu}, courtesy of J.T. Hoeksema. We acknowledge the use of CHIMERA and CHARM databases, which are using SDO/AIA, SDO/HMI, STEREO, SOHO and \textit{Hinode} data sets. We thank Noel Hallemans from the Vrije Universiteit Brussel, for providing the programs to build synoptic maps. We acknowledge support from the Belgian Federal Science Policy Office (BELSPO) through the ESA-PRODEX programme, grant No. 4000120800. We thank the anonymous reviewer for the very useful comments which greatly improved the manuscript. Many thanks to Sarah Willems and Bogdan Nicula for helping with the IDL issues in these times of confinement.
\end{acks}

\section*{Disclosure of Potential Conflicts of Interest}
The authors declare that they have no conflicts of interest. 

\appendix

\section{Synoptic Maps} \label{S-synoptic}

\subsection{On-Disk Synoptic Maps} \label{S-ondisk}

\begin{figure}
	\centering
	\includegraphics[width=0.3\textwidth]{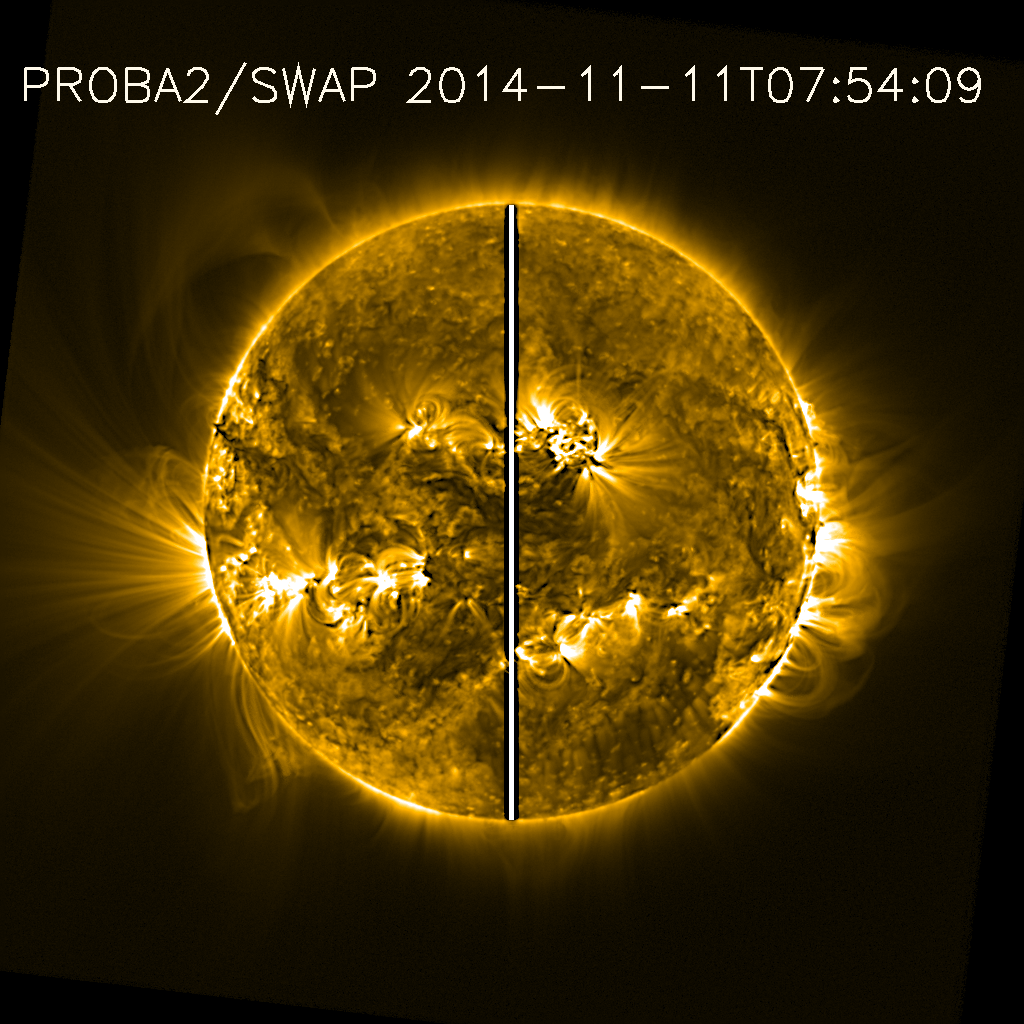}
	\includegraphics[width=0.3\textwidth]{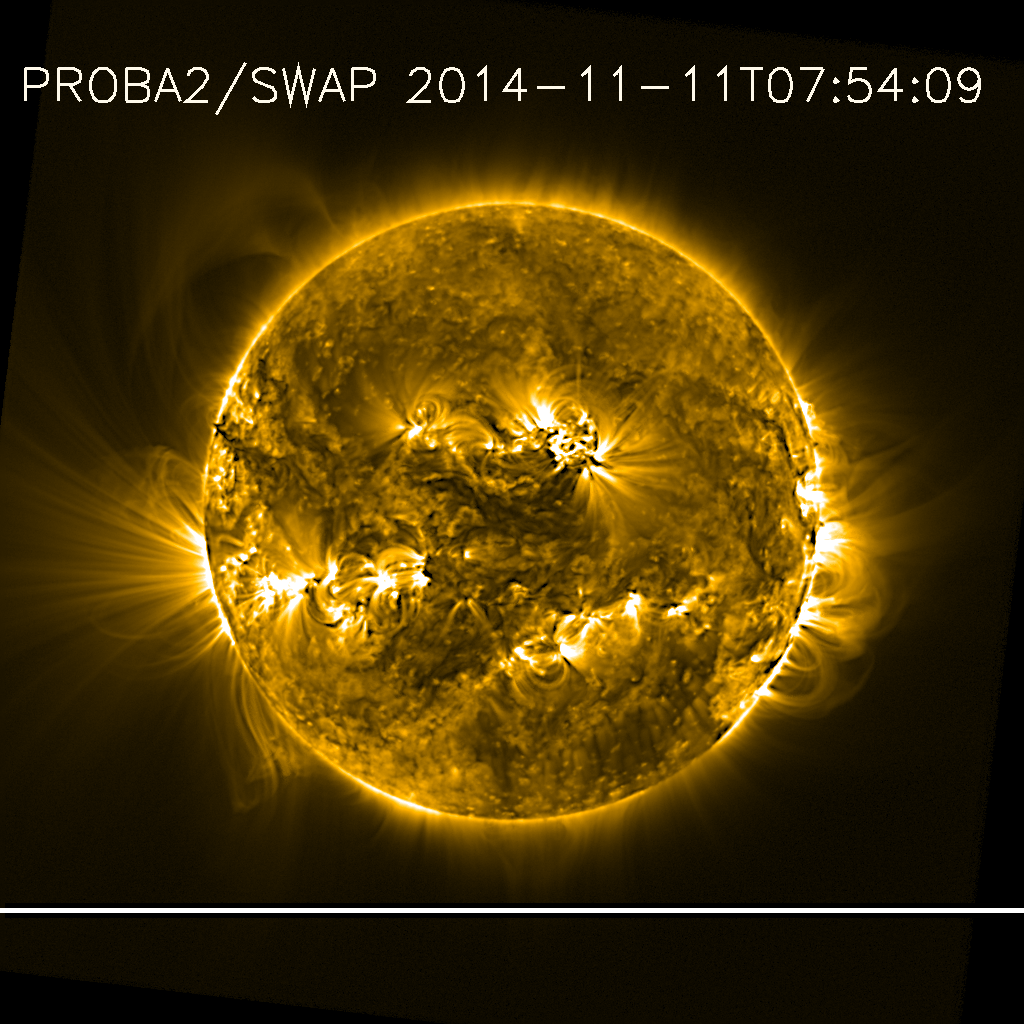}
	\includegraphics[width=0.3\textwidth]{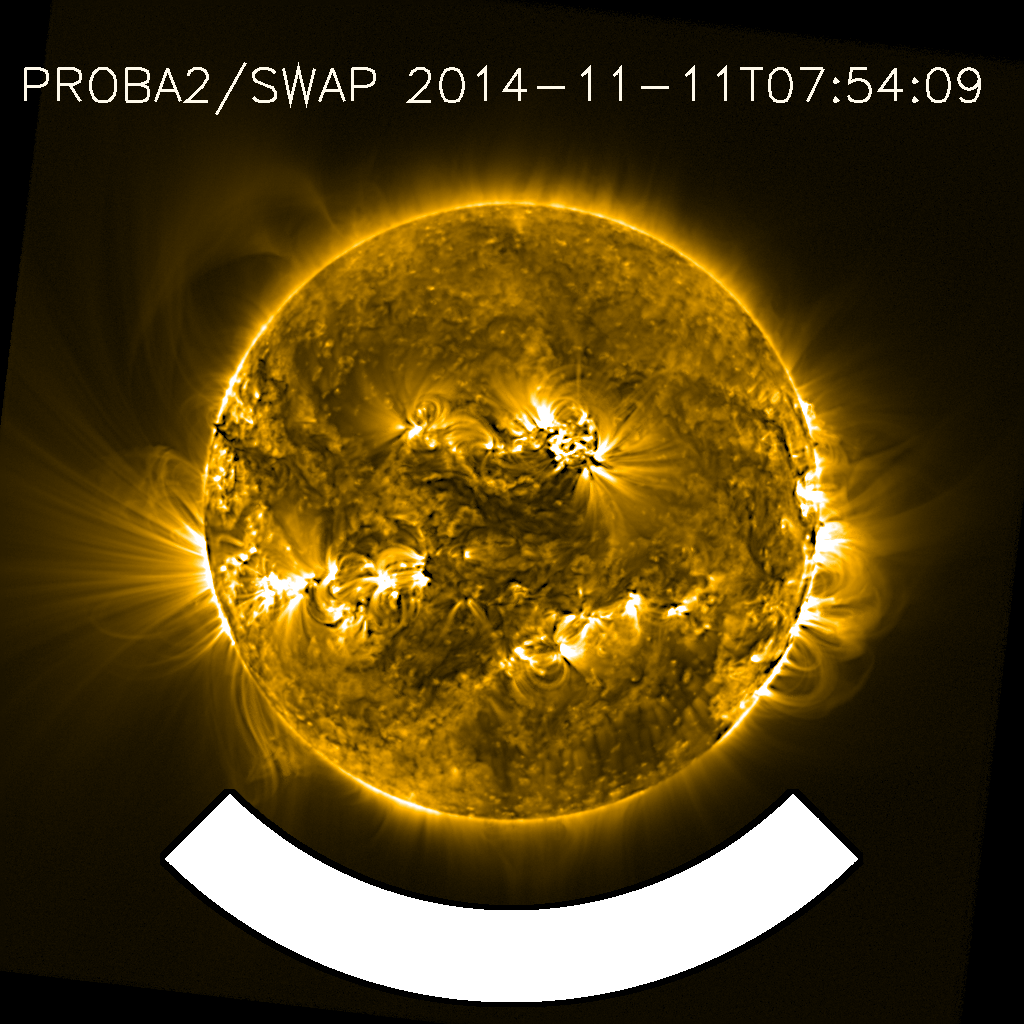}
	\caption{SWAP image on 11 November 2014. \textit{Left panel}: The \textit{vertical white stripe} shows the region at central meridian that is used to build the synoptic map. \textit{Middle panel}: The \textit{horizontal white stripe} shows the region at 1.3 solar radii from the Sun center that is used to build the off-limb synoptic map at South Pole. \textit{Right panel}: Example of a sector where the average value is calculated (\textit{white pixels}). The sector is centered at South Pole, it has a width of 90$^{\circ}$ and the height from 1.3 to 1.6\,R$_{\odot}$.} \label{F-synopticonoff}
\end{figure}

The SWAP on-disk synoptic maps (see, \textit{e.g.}, Figure~\ref{F-swapsynoptic}) were constructed using averaged 3$^{\circ}$-wide longitudinal stripes (see left panel of Figure~\ref{F-synopticonoff}), centered on the central meridian for each image in a CR stack. The synoptic map shows an overview of the Sun for each Carrington rotation as observed by PROBA2/SWAP instrument. The horizontal axis represents the time (in days) and the vertical axis represents the Stonyhurst latitude in degrees. For a description of this coordinate system see \cite{Thompson2006}.

We also build the East--West (EW) SWAP synoptic maps (or time--longitude maps), similar to the classical synoptic maps described above, but this time we used averaged 3$^{\circ}$-wide latitudinal stripes centered on the solar Equator (see Figure~\ref{F-ewswapsynopticlat}). The vertical axis represents the time [days] and the horizontal axis represents the Stonyhurst longitude [degrees].
Similar to this, we built EW synoptic maps at different latitudes ($\pm$20$^{\circ}$, $\pm$40$^{\circ}$) -- see e.g. Figure~\ref{F-ewswapsynoptic2157}. 

\subsection{Off-Limb Synoptic Maps} \label{S-offlimb}

We build off-limb synoptic maps at the Equator and Poles by taking horizontal three-pixels stripes respectively, at 1.1, 1.3, and 1.6 solar radii from the Sun center, average them over the three-pixel width and stack them in time -- see Figure~\ref{F-offlimbimg} for examples of off-limb synoptic maps. In the case of equatorial off-limb maps the time [days] is on the horizontal axis and the helioprojective latitude [arcseconds] is on the vertical axis. The cuts are done at different distances from the Sun center. For the polar off-limb images, the $y$-axis represents the time [days] and the $x$-axis represents the helioprojective longitude [arcseconds]. The central panel of Figure~\ref{F-synopticonoff} shows an example of a horizontal stripe at a distance of 1.3\,R$_{\odot}$ south from the Sun center from which the polar off-limb synoptic maps were created.

\section{Sectors}\label{S-sectors}

In order to study the evolution of the solar corona at different altitudes, and over different regions, we divided each image to sectors where we measured the mean brightness. The right panel of Figure~\ref{F-synopticonoff} shows an example of a sector centered at the South Pole, with a width of 90$^{\circ}$, and the height from 1.3 to 1.6\,R$_{\odot}$. 

\bibliographystyle{spr-mp-sola}
\bibliography{bibliography}

\end{article} 

\end{document}